\documentclass[a4paper,fleqn]{cas-dc}

\usepackage[numbers]{natbib}
\usepackage{amsmath,amsfonts,amssymb}
\usepackage{algorithmic}
\usepackage{algorithm}
\usepackage{array}
\usepackage{siunitx}
\usepackage{mathtools}
\usepackage{booktabs}
\usepackage{multirow}
\usepackage{placeins}
\usepackage{makecell}
\usepackage{threeparttable}
\usepackage{bm}
\usepackage{microtype}
\usepackage{pdflscape}
\begin{document}
\let\WriteBookmarks\relax
\def\floatpagepagefraction{1}
\def\textpagefraction{.001}

\shorttitle{Learning to Remember: Attentive Reinforcement Learning for Edge Serverless Autoscaling}
\shortauthors{F. Shaikh et~al.}

\tnotetext[1]{This work has been supported by the European Union -- NextGenerationEU under the Italian Ministry of University and Research (MUR) Extended Partnerships grant PE00000001 -- RESTART, and by University of Perugia under project SPAIN.}

\title[mode=title]{Learning to Remember: Attentive Reinforcement Learning for Edge Serverless Autoscaling}
% other possible title: Seeing the Recent Past: Attention-Augmented Reinforcement Learning for Kubernetes Pod Autoscaling
\author[1]{Faraz Shaikh}[type=editor,
                        orcid=0009-0005-7815-8283]
\ead{faraz@dottorandi.unipg.it}
\credit{Conceptualization, Methodology, Software, Investigation, Data curation, Visualization, Writing -- original draft}
\affiliation[1]{organization={Department of Engineering, University of Perugia},
    city={Perugia},
    postcode={06125},
    country={Italy}}

\author[1,2]{Gianluca Reali}
\ead{gianluca.reali@unipg.it}
\credit{Supervision, Writing -- review \& editing, Project administration}
\affiliation[2]{organization={Consorzio Nazionale Interuniversitario per le Telecomunicazioni (CNIT)},
    city={Parma},
    postcode={43124},
    country={Italy}}

\author[1,2]{Mauro Femminella}[type=editor,
                        orcid=0000-0002-6695-5956]
\cormark[1]
\ead{mauro.femminella@unipg.it}
\credit{Conceptualization, Supervision, Funding acquisition, Writing -- review \& editing}

\cortext[cor1]{Corresponding author.}

\begin{abstract}
In edge computing, the stochastic and bursty nature of serverless workloads challenges autonomous resource orchestration. Traditional reactive controllers, such as the Kubernetes Horizontal Pod Autoscaler (HPA), suffer from reaction latency, leading to Service Level Objective (SLO) violations during traffic spikes and resource flapping during ramp-downs. While Deep Reinforcement Learning (DRL) offers a pathway toward proactive management, standard agents suffer from \textit{temporal blindness}, an inability to exploit the recent temporal context in non-Markovian edge environments. To bridge this gap, we propose a stability-aware autoscaling framework unifying short-horizon temporal context and control via an Attention-Enhanced Double-Stacked LSTM architecture integrated within a Proximal Policy Optimization (PPO) agent. Unlike shallow recurrent models, our approach employs a learned attention mechanism that weights recent historical states non-uniformly, suppressing high-frequency jitter while preserving the trend that precedes demand shifts. We validate the framework on two independent Kubernetes clusters using real-world Azure Functions traces. %, with every learning agent evaluated across five random seeds per cluster. 
Against the single-layer LSTM ablation and the static HPA baseline, our approach reduces P90 latency by $\approx$67\%, and holds average latency within the 50ms hard SLO for 98.8\% of the run against 49.6\% and 43.5\% respectively. Against Kubernetes Event-Driven Autoscaling (KEDA), it matches latency performance at 75\% fewer replica-steps and 59\% less churn, with P90 hard-SLO violation bursts of at most 5 consecutive intervals against up to 24 for KEDA. These results indicate that mitigating temporal blindness through deep attentive memory improves the reliability and stability of Kubernetes autoscaling under bursty edge workloads.
\end{abstract}
\begin{comment}
\begin{highlights}
\item A Kubernetes-native supervisory RL controller maps policy actions to HPA CPU targets and gateway rate limits, not raw replica counts.
\item An Attention-Double-LSTM policy cuts P90 tail latency by 67\% over a single-layer LSTM ablation on real Azure traces.
\item Validated on two independent clusters, five seeds each, with matched-pair significance testing (Wilcoxon, $p{=}0.0020$).
\item Matches KEDA's latency with 75\% fewer replica-steps and 59\% less replica churn.
\end{highlights}
\end{comment}

\maketitle

\section{Introduction}
\label{sec:introduction}
The architectural paradigm of modern digital services has undergone a fundamental transformation toward cloud-native, edge, and serverless computing. Driven by the strict low-latency requirements of 5G/6G applications and the Internet of Things (IoT), computational resources are increasingly getting decentralized and positioned toward the network edge \citep{wang2023wireless}. Within this ecosystem, applications are no longer monolithic, and are decomposed into loosely coupled microservices or serverless functions orchestrated by platforms such as Kubernetes \citep{dogani2023auto}. This modularity provides unprecedented agility, allowing individual components to scale independently. Conversely, it introduces an important challenge in autonomous resource orchestration. Unlike traditional data center workloads manifesting predictable diurnal patterns, serverless functions in edge-cloud environments are subject to highly stochastic, non-stationary, and bursty invocation patterns \citep{xu2025auto}. Therefore, service providers face the complex optimization problem of dynamically adjusting resources i.e., \textit{autoscaling}, to adhere to strict Service Level Objectives (SLOs) and Quality of Service (QoS) standards while minimizing operational expenditure (OPEX) and avoiding resource wastage.

\begin{figure*}
    \centering
    \includegraphics[width=\linewidth]{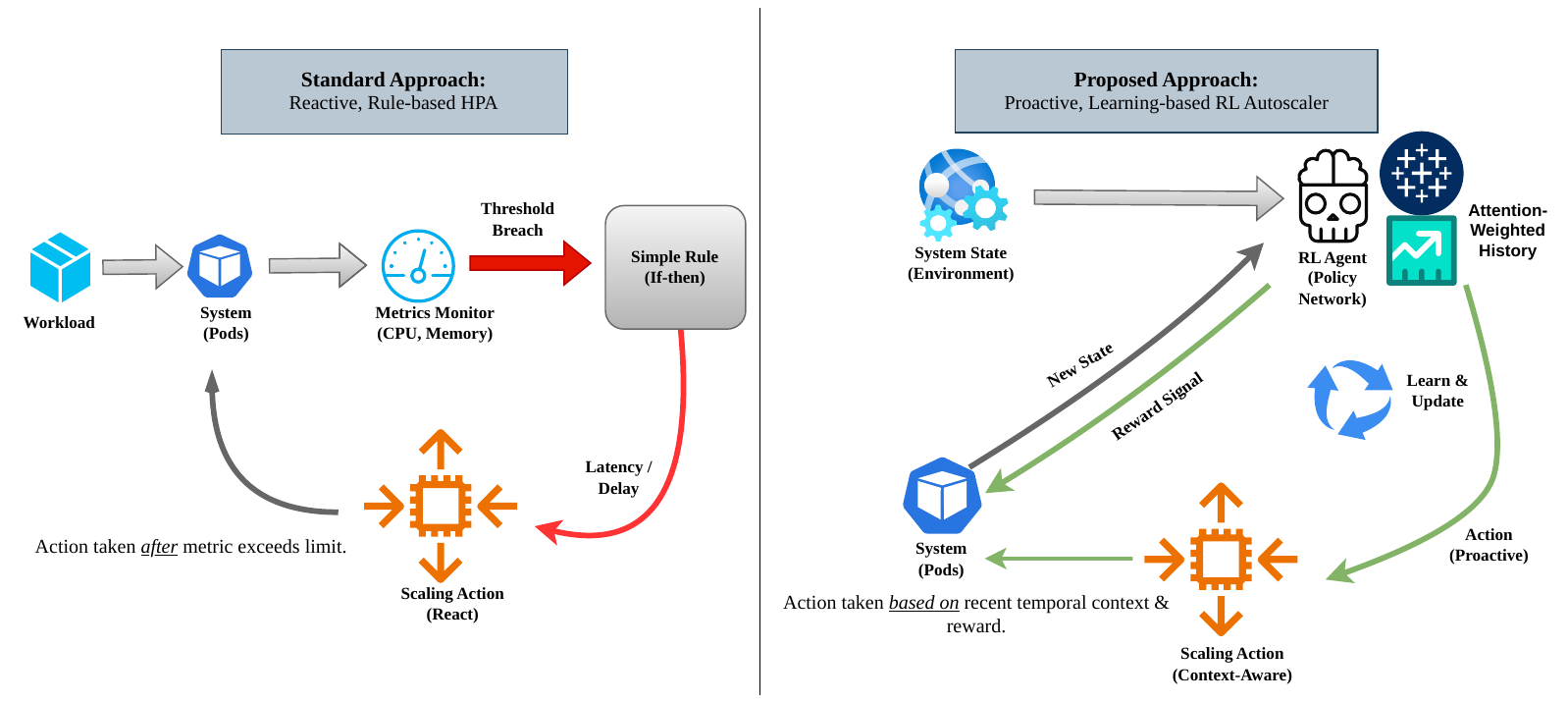}
    \caption{Conceptual comparison of autoscaling paradigms. \textit{(Left)} The standard HPA approach vs. \textit{(Right)} Proposed stability-aware Reinforcement Learning framework. ``Proactive'' here means sizing each response decisively from recent history, not scaling ahead of the demand curve; the agent's replica trajectory still follows the workload (Section~\ref{sec:discussion}).}
    \label{fig:concept_comparison}
\end{figure*}

The industry-standard for horizontal autoscaling, represented by the Kubernetes Horizontal Pod Autoscaler (HPA), operates primarily through reactive feedback-control loops, as illustrated in Fig.~\ref{fig:concept_comparison}, scaling container replicas based on observed resource utilization metrics. These controllers monitor aggregate metrics, such as CPU and/or memory utilization, and trigger scaling actions only when a threshold is violated for a specific duration. While robust in steady-state scenarios, reactive techniques suffer from a fundamental \textit{reaction latency}, i.e., the unavoidable delay between the onset of a traffic surge and the eventual readiness of new function replicas \citep{golec2024cold}. This delay is further complemented by container cold starts and initialization overheads, leading to periods of under-provisioning where SLO violations are unavoidable. Furthermore, to prevent \textit{flapping} (i.e., rapid oscillation), reactive controllers often employ \textit{hysteresis}, also known as cooling periods, leading to resource misalignment by either retaining idle resources during ramp-downs or by failing to react quickly enough to successive bursts \citep{k8s_hpa_flapping}.

Beyond the temporal limitations of reaction latency, heuristic and threshold-based approaches fundamentally struggle to capture the complex, non-linear mapping between infrastructure metrics and application performance \citep{rossi2020hierarchical}. In heterogeneous edge environments, the correlation between resource utilization and user-perceived latency is hardly static, fluctuating dynamically based on request complexity, downstream service dependencies, and background interference \citep{khaleq2021intelligent}. Hence, a static threshold that ensures SLO compliance during one operational window may result in severe violations or resource wastage during another, as infrastructure metrics often fail to serve as accurate proxies for application-level QoS. This variability renders the manual tuning of scaling policies complex at scale, as human operators cannot continuously adjust parameters to match the evolving system dynamics \citep{xu2025auto}.

To cope with these challenges, researchers have found Deep Reinforcement Learning (DRL) as a viable strategy for autonomous service management. DRL agents can learn optimal scaling strategies by interacting with system environment, making real-time decisions on when to retain, increase, or decrease the number of instances (replicas) to handle fluctuating traffic loads effectively \citep{xiao2022dscaler}. Early adoption involved value-based methods such as Deep Q-Networks (DQN) \citep{kim2022improved,lee2021deep}, which demonstrated improvements over static thresholds in optimizing energy and execution time. More recently, policy-gradient methods, such as Proximal Policy Optimization (PPO) have been applied to serverless edge computing, marking superior stability in continuous control tasks \citep{santos2025can}. A critical limitation of these approaches lies in their failure to address the non-Markovian nature of edge workloads \citep{cortez2017resource,hausknecht2015deep}. While functions as a service (FaaS) architectures simplify scaling by decoupling computation from state, standard DRL agents fail to capture the historical context required for accurate forecasting \citep{hernandez2019survey}. By operating on a single-step observation window, these agents suffer from partial observability, manifesting as \textit{temporal blindness}, a failure to differentiate between momentary jitter and genuine shifts in traffic intensity.

Recent efforts have tried to mitigate this short-coming by integrating Recurrent Neural Networks (RNNs) or Long Short-Term Memory (LSTM) units. However, existing frameworks are predominantly based on shallow, single-layer architectures \citep{agarwal2024deep}. These standard recurrent models lack the representational depth to capture multi-scale dependencies, such as differentiating short-lived outliers from the onset of sustained shift in demand \citep{zhao2025mscnet}, and process historical sequences with uniform weight. Hence, they cannot preferentially weight the most informative recent intervals, such as the rapid initiation of a concurrent user spike, while filtering out irrelevant background noise \citep{meng2023deepscaler}.

Even in LSTM-PPO autoscalers such as DRe-SCale \citep{agarwal2024deep}, (i) shallow recurrence still exhibits an information bottleneck, (ii) the control interface is often replica-centric rather than Kubernetes knob-centric, and (iii) safe translation from policy actions to operational configurations is typically not explicit.

In light of these limitations, we propose a novel, stability-aware autoscaling framework unifying temporal context and control via an Attention-enhanced Double-Stacked LSTM agent, a concept of which is also illustrated in Fig.~\ref{fig:arch}. Unlike prior works that treat workload prediction as a separate module feeding a controller, the proposed agent performs temporal inference inside the policy network itself, conditioning each scaling decision on a learned representation of recent history rather than on an externally supplied forecast. We employ a double-stacked LSTM architecture to model high-order traffic dependencies, capturing both immediate fluctuations and multi-step trend, and a learned attention mechanism to weigh the significance of recent historical states non-uniformly. As a result, the agent constructs a context-aware representation of the recent workload, suppressing stochastic single-step jitter while preserving the short-horizon trend that precedes demand shifts. Furthermore, to handle the operational complexity of production environments, our agent operates in a multi-dimensional discrete action space, simultaneously optimizing HPA CPU targets and throughput multipliers.

Central to this design is an explicit partial-observability argument. Rather than engineering an observation vector intended to approximately restore the Markov property, which is the standard treatment in DRL autoscaling, we formulate the problem as a partially observable Markov Decision Process (POMDP) in which queue depths and instantaneous arrival processes remain hidden by construction. Section~\ref{obs} establishes that no single-step observation of a Kubernetes cluster can be Markovian under bursty edge workloads, and Section~\ref{arch} derives the recurrent policy as the direct consequence of that fact. Recurrence here therefore follows directly from this problem formulation rather than being added empirically to an otherwise stateless agent. What separates the proposed framework from prior LSTM-augmented autoscalers, detailed further in Section~\ref{sec:literature}, is the combination of recurrent depth, attention, and a Kubernetes-native supervisory control interface, rather than the presence of recurrence alone.

The scientific contributions of this article are threefold:
\begin{enumerate}
    \item We introduce an Attention-enhanced Double-Stacked LSTM policy network that specifically addresses the \textit{temporal blindness} of stateless DRL and improves over single-layer LSTM-PPO policies by attending to critical historical timesteps while filtering transient noise.
    \item We provide a formal, comparative benchmarking against Kubernetes Event-Driven Autoscaling (KEDA), a Single-LSTM PPO variant, and a standard rule-based HPA explicitly isolating the gain from (i) stacking depth and (ii) attention under non-Markovian workloads evaluated over five random seeds on two heterogeneous clusters rather than a single run on a single testbed. Complementary single-factor ablations isolate the contribution of the attention context width and of the auxiliary control actuator, separating the effect of temporal context from that of recurrent capacity.
    \item We conduct an extended stability analysis using real-world Azure Functions invocation traces. Beyond standard metrics, such as latency and CPU usage, we quantify operational stability using replica churn and oscillation frequency, demonstrating that our proposed framework reduces unnecessary scaling actions by a significant margin compared to reactive HPA and standard DRL baselines and show that scaling activity alone is not a valid stability criterion.
\end{enumerate}

This manuscript extends the preliminary conference version \citep{11297411}, revising the controller design (action interface and reward definition) and adopting a disjoint multi-day train/test evaluation protocol. In addition, the overall setup has been tested on a different and more performing server, equipped with a GPU. These changes alter both the optimized objective and the workload exposure, so numerical results are expected to significantly differ.

The remainder of the paper is arranged as follows. Section~\ref{sec:literature} discusses the state-of-the art in this specific research domain. Section~\ref{sec:methodology} details the methodology used for this approach. Section~\ref{results} discusses the results of the test runs conducted. Finally, Section~\ref{sec:discussion} concludes the paper with a discussion on the proposed approach, its current limitations, and potential avenues for future research.

\section{Related Work}
\label{sec:literature}
The evolution of autoscaling in cloud-native environments has transitioned from reactive, rule-based heuristics to sophisticated data-driven controllers. This progression is differentiated by a shift from mathematical stability models to model-free reinforcement learning (RL), and most recently, to hybrid architectures that attempt to solve the ``temporal blindness'' of pure RL agents. This section categorizes the state-of-the-art into reactive control, pure RL, and hybrid recurrent architectures.

The earliest generation of autoscaling systems focused on establishing mathematical stability through feedback control. The industry-standard Kubernetes HPA operates on a reactive, threshold-based loop, monitoring CPU/memory metrics at discrete intervals (typically 15 seconds) and triggering scaling only when thresholds are violated \citep{kubernetes_hpa,francois2025kubernetes}. In steady-state scenarios with predictable workloads, HPA maintains stable performance. However, under bursty traffic, utilization-based scaling lags behind workload changes result in transient overload and SLO violations \citep{punniyamoorthy2025slo,tran2024optimized}. This latency is intensified by container startup times averaging 259\,ms and reaching up to 329\,ms augment the delay between detection and mitigation \citep{hall2022opportunities}. To address these limitations, hierarchical approaches, such as \textit{Gwydion} \citep{santos2025gwydion}, decouple application goals from infrastructure metrics, achieving approximately 28\% improvement in multi-tier topologies. However, the problem often lies in assuming homogeneous infrastructure incompatible with the non-stationary nature of edge-cloud environments \citep{park2023fully}.

Further, to overcome the rigidity of static thresholds, early research pivoted toward value-based RL methods capable of learning non-linear dynamics. Lee et al.~\citep{lee2021deep} applied Deep Q-Network (DQN) to manage server instance scaling in multi-access edge computing (MEC) environments. Their approach demonstrated that learning-based agents could outperform static thresholds in optimizing energy and execution time by adapting to different workload phases. Further, Benedetti et al.~\citep{BENEDETTI2026108112} explored Q-Learning for edge autoscaling, offering a computationally lightweight solution for resource-constrained nodes. However, these value-based systems typically rely on low-dimensional state representations and discrete action spaces. This limits their applicability to containerized microservices/serverless platforms where continuous control and precise resource percentages are required. Furthermore, Q-Learning often demonstrates instability in the high-variance, bursty traffic patterns typical of production environments \citep{BENEDETTI2026108112}.

Recognizing the limitations of Q-Learning in continuous action spaces, recent efforts have shifted toward policy-gradient methods, specifically PPO. Gan et al.~\citep{gan2022adaptive} proposed a PPO framework for edge computing achieving an 86\% improvement over Q-Learning by stabilizing the policy update process through clipping. Addressing specific workload types, the KIS-S framework~\citep{zhang2025kis} utilized PPO for GPU inference autoscaling, bridging the gap between simulation and real hardware. To improve temporal awareness without recurrent architectures, Femminella and Reali~\citep{femminella2024application} integrated cyclic time-of-day encoding with a PPO agent. While this allowed the agent to differentiate between peak and off-peak hours, the agent remained fundamentally reactive, lacking an internal memory mechanism to estimate future demand based on historical sequences. Alternative approaches have attempted to operate at the OS level. FaaSCtrl~\citep{panda2024faasctrl} used an Advantage Actor-Critic (A2C) controller to manage tail latency by tuning Linux scheduling parameters. However, by operating at the OS level rather than the orchestration level, these solutions lack direct portability to standard Kubernetes deployments.

%Multi-objective resource optimization in heterogeneous environments has also received attention. Joint optimization of throughput, energy, and QoS in UAV-enabled heterogeneous cellular networks using serverless federated learning and power-domain NOMA~\citep{noma2025} highlights the breadth of resource management challenges that motivate multi-objective reward formulations. 
The current frontier of research addresses the non-Markovian nature of edge workloads by integrating time-series forecasting with control loops. Initial attempts focused on decoupled architectures, known as \textit{Separation of Concerns}. Peng et al.~\citep{peng2023microservice} and Yan et al.~\citep{yan2021hansel} used Bi-LSTM networks for workload prediction, achieving high accuracy of up to 92.3\%. However, decoupling prediction from control often leads to error propagation, where small forecast errors trigger cascading scaling oscillations \citep{kim2024self,lim2021temporal}. Gupta et al.~\citep{gupta2025hybrid} combined proactive Machine Learning (ML)-based scaling with reactive safety nets, reducing SLO violations to 6\%. However, the cons for this approach included a significant increase in configuration complexity and potential conflicts between the dual control loops.

More recently, the research has pivoted more towards the lightweight predictive models and heavy-duty RL agents. On the predictive front, Guruge and Priyadarshana~\citep{guruge2025time} proposed a hybrid forecasting framework combining Facebook Prophet with LSTM networks, especially aiming at resolving the cold-start latency of standard HPA by predicting HTTP request patterns ahead of time, addressing the complexity of multi-objective decision-making without the training overhead of ML. Further, novel mathematical formulations, such as \textit{qAHP}~\citep{dimolitsas2026enabling} have been introduced to enable real-time autoscaling in edge clusters, reducing decision latencies by orders of magnitude compared to traditional analytic hierarchy processes.

To unify prediction and control, researchers are increasingly embedding memory directly into the RL policy. Ma et al.~\citep{ma2024auto} proposed combining Fuzzy Q-Learning with LSTM networks, though the fuzzy logic introduced computational overheads unsuitable for the edge. The tightest integration to date is \textit{DRe-SCale} by Agarwal et al.~\citep{agarwal2024deep}, which embeds an LSTM layer directly within a PPO policy. In serverless environments, this LSTM-PPO architecture improved throughput by 18\% compared to non-recurrent baselines by addressing partial observability.

Despite these advancements, a critical architectural gap remains. Current hybrid approaches, such as \textit{DRe-SCale}, rely on shallow, single-layer LSTM architectures that suffer from the information bottleneck, where the entire history is compressed into a fixed-size hidden state. Consequently, these approaches struggle to separate signal from noise within their recent context, often failing to selectively attend to critical precursors, such as the onset of a micro-burst, masked by background fluctuations \citep{kim2024self}. Furthermore, graph-attentive multi-agent reinforcement learning (MARL) for joint SLA-aware task offloading and adaptive service orchestration~\citep{11430686} illustrates the potential of advancing structural service dependencies, informing the multi-agent extension we identify as future work. Our work addresses this limitation by introducing an Attention-enhanced Double-Stacked LSTM policy. By integrating deep temporal attention, our model retains the sequential context of LSTMs while gaining the ability to explicitly weigh high-importance historical events, thereby unifying temporal context and control.

\section{Methodology and System Architecture}
\label{sec:methodology}
This study reformulates the challenge of Kubernetes autoscaling as a sequential decision-making problem under uncertainty, solved via a DRL approach. We present a comprehensive control framework integrating short-term temporal context with a multi-objective policy network. To validate the proposed architecture, we establish a comparative experimental design involving four different scaling strategies, evaluated across two independent heterogeneous hardware clusters.

While the constituent components of our framework, i.e., LSTM, PPO, and soft-attention, are individually well-established, our contribution lies in their integration as a supervisory controller within the Kubernetes control plane. Rather than outputting raw replica counts, our state-to-action mapping (Section~\ref{action_space}) produces Kubernetes-native configuration parameters (HPA utilization targets and gateway rate limits), preserving Kubernetes' own fail-safe mechanisms while layering learned adjustments on top of them. This is coupled with a reward (Section~\ref{reward}) that explicitly bridges application-level QoS with infrastructure-level state, preventing the degenerate provisioning behavior common to standard RL formulations.

\subsection{Problem Formulation and Observation Space ($\mathcal{S}$)}
\label{obs}
We formally model the autoscaling domain as a POMDP. Unlike a standard Markov Decision Process (MDP) where the agent has perfect knowledge of the environment, a POMDP assumes the agent only observes a partial representation of the true system state \citep{agarwal2024deep}. In a Kubernetes cluster, factors such as network queue depths and instantaneous request arrival processes are often hidden or noisy. Therefore, the agent must rely on a constructed observation vector $\mathbf{s}_t$ to infer the system's true health and make optimal scaling decisions.

To resolve the partial observability characteristic of bursty edge workloads, the agent constructs a comprehensive 13-dimensional state vector $\mathbf{s}_t$. This vector aggregates immediate infrastructure health with a short-horizon demand estimate into a compact observation. Because network queue depths and request arrival processes introduce residual temporal correlations that no single-step snapshot can fully capture, an individual $\mathbf{s}_t$ deliberately does \emph{not} satisfy the Markov property. This is precisely the motivation for the recurrent architecture described in Section~\ref{arch}. Rather than acting on a single frame, the policy consumes the $w_a = 8$ most recent observation vectors as one $8 \times 13$ input, encodes them with a stacked LSTM and weighs them by attention, so that the temporal context an individual frame omits is recovered from the window itself. The state is formally defined as the concatenation of five feature groups, in Eq.~\eqref{eq:obs_space}:
\begin{equation}
    \mathbf{s}_t = \left[ \mathbf{m}^{perf}_t, \mathbf{m}^{res}_t, \mathbf{m}^{conf}_t, \mathbf{x}^{time}_t, \hat{N}_t \right]
    \label{eq:obs_space}
\end{equation}
\noindent where:
\begin{itemize}
    \item $\mathbf{m}^{perf}_t \in \mathbb{R}^3$: Captures the Quality of Service (QoS) and load intensity, specifically the average request latency $l_t$, the success ratio $SR_t$, and the \textit{effective} incoming request rate $\lambda_t$ (i.e., the rate of requests successfully admitted by the API gateway after filtering).
    \item $\mathbf{m}^{res}_t \in \mathbb{R}^5$: Represents the resource saturation levels, including the current replica count $\rho_t$, average pod utilization ($u^{cpu}_t, u^{ram}_t$), and total cluster-wide resource consumption ($U^{total}_{cpu}, U^{total}_{ram}$). Integrating cluster-wide metrics allows the agent to sense node-level saturation risks. Although $u^{ram}_t$ and $U^{total}_{ram}$ inform the agent's perception of memory pressure, the reward function (Section~\ref{reward}) carries no memory term: the evaluated \texttt{factorizator} workload is CPU-bound and RAM utilization stays far from saturation throughout all experiments. Extending the reward to memory- or I/O-bound services, where RAM would need to enter the objective directly, is discussed as future work in Section~\ref{sec:discussion}.
    \item $m_t^{conf} \in \mathbb{R}^2$: Encodes the active control configuration to provide context on previous decisions, comprising only the current HPA CPU target and the throughput multiplier.
    %\item $\mathbf{x}^{time}_t \in \mathbb{R}^2$: Cyclic temporal embeddings $[\cos(\frac{2\pi \tau_t}{T_{day}}), \\\sin(\frac{2\pi \tau_t}{T_{day}})]$, where $\tau_t$ denotes the clock-time-of-day in minutes at step $t$ (as opposed to the control-step index itself), allowing the policy to encode diurnal traffic periodicities ($T_{day}=1440$ min).
    \item $\mathbf{x}^{time}_t \in \mathbb{R}^2$: Cyclic temporal embeddings $[\cos(\frac{2\pi \tau_t}{T_{day}}), \\\sin(\frac{2\pi \tau_t}{T_{day}})]$, where $\tau_t$ is the control-step index within the current simulated day and $T_{day}=500$ control intervals, the length of one sampled trace-day at the 60\,s control interval (Section~\ref{testbed-description}), allowing the policy to encode within-day periodicity rather than a 24-hour diurnal cycle.
    \item $\hat{N}_t \in \mathbb{R}^1$: A short-horizon demand estimate computed from a sliding window over recent requests (window $w_f=3$) with exponential smoothing to stabilize the signal used by the controller.
\end{itemize}

\subsection{Multi-Dimensional Discrete Action Space ($\mathcal{A}$)}
\label{action_space}
The agent controls the cluster via a Multi-Discrete action space $\mathcal{A} \in \mathbb{Z}^2_{\geq 0}$, allowing simultaneous manipulation of two operational parameters at each time step $t$. The action vector is defined as $\mathbf{a}_t = \left[ a^{targ}_t, a^{mult}_t \right]$, where:
\begin{itemize}
    \item HPA CPU Target ($a^{targ} \in \{0, 1, 2, 3\}$): Dynamically adjusts the target utilization setpoint for the underlying HPA. This is mapped to the discrete set $\{30\%, 50\%, 70\%, 90\%\}$, allowing the agent to switch strategies between conservative resource buffering (low target) and aggressive consolidation (high target) based on workload volatility.
    \item Throughput Multiplier ($a^{\text{mult}} \in \{0, 1, 2\}$): Scales the API gateway's admitted request rate to $\{1.0\times, 2.0\times, 3.0\times\}$ the rate specified by the replayed trace. Because $1.0\times$ is its lowest setting, this action lets the agent decline to amplify demand during overload. It cannot admit less than the trace's baseline rate. Selecting $1.0\times$ under overload is therefore a refusal to amplify rather than active load shedding.
\end{itemize}
At each step, discrete action indices are mapped deterministically to valid configuration values (HPA target and gateway multiplier) and applied via Kubernetes API updates. Actions are clipped to deployment bounds, such as target setpoints and multiplier ranges, to ensure feasibility.

\begin{figure*}
    \centering
    \includegraphics[width=\linewidth]{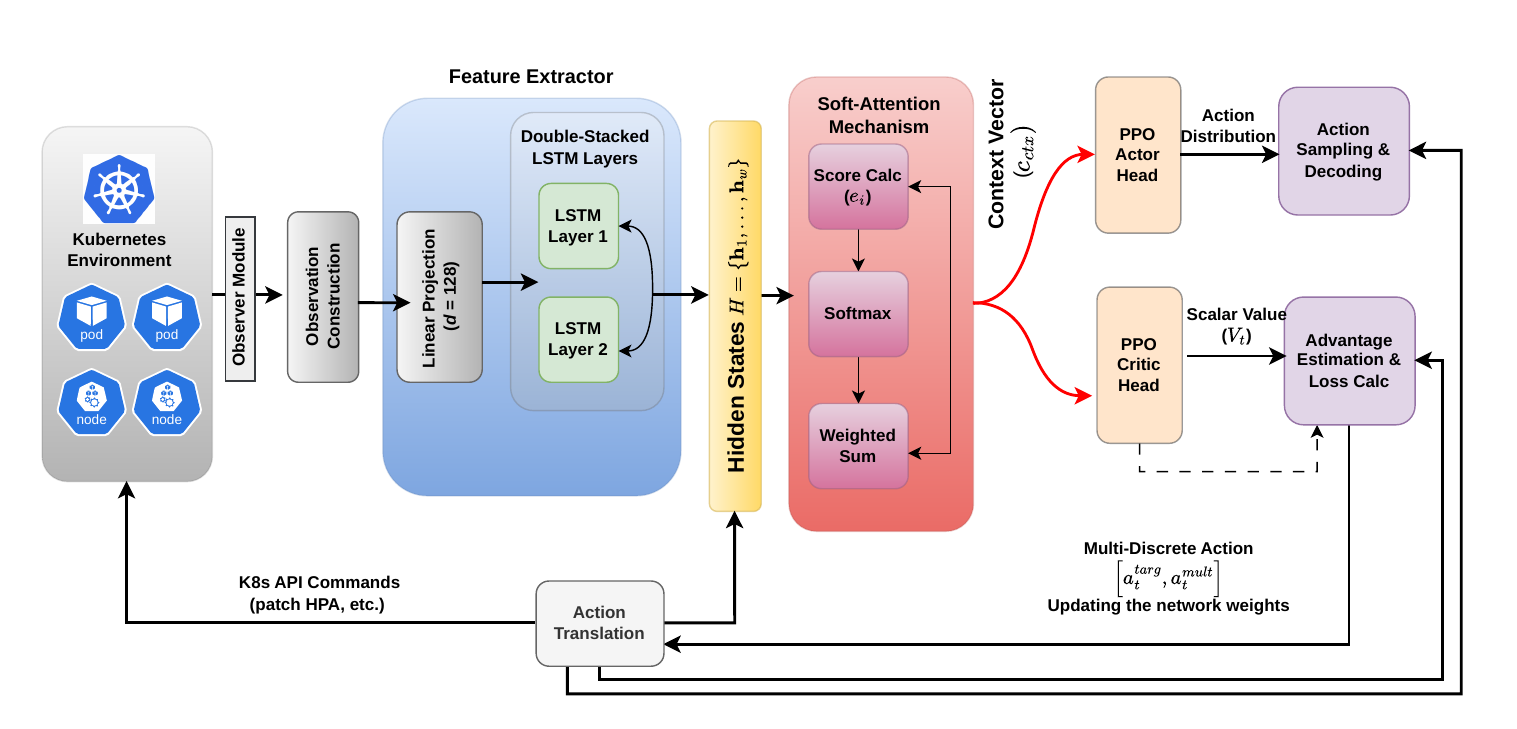}
    \caption{Attention-enhanced Double-Stacked LSTM--PPO autoscaler: embeds Kubernetes observations, applies soft attention over stacked LSTMs for actor--critic action selection, and translates actions into Kubernetes API commands with PPO updates.}
    \label{fig:arch}
\end{figure*}

\subsection{Optimization Objective}
We employ PPO, an on-policy gradient method, for its stability. To prevent destructive policy updates during volatile training phases, we augment the standard clipped objective with a Kullback-Leibler (KL) penalty, given in Eq.~\eqref{eq:KL}.
\begin{equation}
\resizebox{1\columnwidth}{!}{$
    L(\theta) = \mathbb{E}_t \left[ \min\left(r_t(\theta)\hat{A}_t, \text{clip}(r_t(\theta), 1-\epsilon, 1+\epsilon)\hat{A}_t\right) - \beta \text{KL}(\pi_{\theta_{old}}, \pi_\theta) \right]
$}
\label{eq:KL}
\end{equation}
where $\epsilon=0.2$ is the clipping range and $\beta$ dynamically scales the KL penalty. Here, $r_t(\theta) = \tfrac{\pi_\theta(\mathbf{a}_t|\mathbf{s}_t)}{\pi_{\theta_{old}}(\mathbf{a}_t|\mathbf{s}_t)}$ denotes the probability ratio of the action under the current and previous policies, and $\hat{A}_t$ represents the generalized advantage estimate \citep{schulman2017proximal}, which measures how better an action is compared to the expected baseline. This formulation ensures the new policy $\pi_\theta$ does not deviate excessively from the behavioral policy $\pi_{\theta_{old}}$.

\subsection{SLO-Aware Reward Engineering}
\label{reward}
The reward function $R_t$ is a composite scalar designed to guide the agent toward a Pareto-optimal balance between strict SLO compliance and resource efficiency. It is computed as a weighted sum, shown in Eq.~\eqref{eq:reward}. The weights were selected through iterative empirical tuning guided by explicit design priorities. SLO compliance ($w_{\text{slo}} = 0.55$) receives the dominant weight, as latency guarantees are the primary objective. Resource efficiency ($w_{\text{cpu}} = 0.25$) prevents chronic over-provisioning, and the success ratio ($w_{\text{succ}} = 0.12$), and stability ($w_{\text{stab}} = 0.08$) terms limit unreliable and oscillatory behavior. To validate robustness, we conducted a post-hoc sensitivity analysis in which each weight was changed by $\pm 50\%$ and the mean reward of every agent recomputed on the reported evaluation runs. The ranking of agents was preserved in all twelve resulting configurations, and also across all $81$ combinations of the joint sweep in which the four weights were varied simultaneously, where the smallest margin separating the proposed agent from the next-best baseline is $0.067$. The reward responds mainly to the two dominant terms: over the changing range, the SLO weight moves the proposed agent's mean reward by $0.449$ and the CPU-efficiency weight by $0.230$, against $0.120$ for the success ratio and $0.055$ for stability.
\begin{equation}
    R_t = w_{slo} R_{SLO} + w_{cpu} R_{CPU} + w_{succ} R_{Succ} + w_{stab} R_{Stab}
    \label{eq:reward}
\end{equation}
\noindent where the components are defined as follows:

\begin{itemize}
    \item SLO Compliance ($R_{SLO}$): To enforce low-latency guarantees, we define a target latency $L_{target}=20$ms as the desired \textit{good QoS} operating target and a hard violation threshold $L_{thresh}=50$ms to create a two-tier latency objective that rewards staying fast while strongly penalizing overload. The reward is formulated as a piecewise function, given in Eq.~\eqref{eq:slo}:
    \begin{equation}
    \resizebox{0.9\columnwidth}{!}{$
        R_{SLO} =
        \begin{cases}
        1.0 & \text{if } l_t \le L_{target} \\
        0.5 + 0.5 \frac{L_{thresh} - l_t}{L_{thresh} - L_{target}} & \text{if } L_{target} < l_t \le L_{thresh} \\
        \max(-1.0, -0.5 \frac{l_t - L_{thresh}}{0.1}) & \text{if } l_t > L_{thresh}
        \end{cases}
    $}\label{eq:slo}
    \end{equation}
    This structure provides a maximum reward for meeting the target, a linear decay for \textit{graceful degradation} within the threshold, and a sharp negative penalty for SLO violations.

    \item Resource Efficiency ($R_{CPU}$): To prevent over-provisioning, we model efficiency as a Gaussian function centered at the agent's chosen HPA target $T_{hpa}$, given in Eq.~\eqref{eq:r_cpu}:
    \begin{equation}
    \resizebox{0.9\columnwidth}{!}{$
        R_{CPU} = \begin{cases}
        1.0 & \text{if } |u^{cpu}_t - T_{hpa}| \le 10 \\
        \exp\left( - \left(\frac{u^{cpu}_t - T_{hpa}}{50}\right)^2 \right) & \text{otherwise}
        \end{cases}
    $}
    \label{eq:r_cpu}
    \end{equation}
    This incentivizes the agent to align the actual cluster usage $u^{cpu}_t$ with its selected target $T_{hpa}$, inclusive of a $\pm10\%$ tolerance buffer to prevent micro-adjustments. This term rewards the agent for internal consistency by selecting a target and then achieving it, rather than proximity to a globally optimal utilization level, which is a deliberate design choice to stabilize the joint optimization of HPA target selection and replica provisioning.

    \item Stability ($R_{Stab}$): A penalty term proportional to the magnitude of scaling actions ($|\Delta \rho_t|$) to discourage control oscillation (flapping). The penalty is formulated in Eq.~\eqref{eq:r_stab}: small changes ($|\Delta \rho_t| \le 2$) incur a minor penalty, while larger jumps are penalized more heavily to promote smooth transitions.
    \begin{equation}
    R_{Stab} =
    \begin{cases}
    -0.1 |\Delta \rho_t| & \text{if } |\Delta \rho_t| \le 2 \\
    -0.5 |\Delta \rho_t| & \text{if } |\Delta \rho_t| > 2
    \end{cases}
    \label{eq:r_stab}
    \end{equation}

    \item Success Ratio ($R_{Succ}$): To ensure the agent maintains high service reliability, we reward the successful completion of requests ($SR_t$) as shown in Eq.~\eqref{eq:r_succ}:
    \begin{equation}
    R_{Succ} =
    \begin{cases}
    1.0 & \text{if } SR_t \ge 0.99 \\
    \log(SR_t) & \text{otherwise}
    \end{cases}
    \label{eq:r_succ}
    \end{equation}
    $SR_t$ is the fraction of the requests \texttt{hey} recorded in the control interval that returned HTTP 200; a request counts as a failure whether it returns a non-2xx status or exceeds \texttt{hey}'s default 20\,s per-request timeout.
\end{itemize}

To clarify the design rationale, centering the $R_{CPU}$ reward on the agent's selected HPA target $T_{hpa}$ ensures operational consistency. Rather than chasing a static, universally optimal CPU utilization, which often introduces thrashing between the agent and the underlying Kubernetes PID loop, this formulation forces the agent to accurately provision resources that sustain its explicitly declared Kubernetes target.

The demand estimate $\hat{N}_t$ enters through the observation vector rather than the reward. A forecast-alignment penalty would encode a specific control strategy in the objective and, not being potential-based, would shift the optimum away from the latency and efficiency criteria, penalizing the headroom required during demand ramps. The choice is supported post-hoc. The learned policy attains an alignment penalty $-((\rho_t C - \hat{N}_t)/\hat{N}_t)^2$ of $-0.166$ (C1) and $-0.183$ (C2), against $-0.304$ for KEDA and $-0.548$ for the Single-LSTM ablation (paired difference $-0.355$, 95\% CI $[-0.731, -0.118]$, identical traces). Capacity tracking therefore emerges from the primary objective. Since any smooth trajectory aligns better with a smoothed target, the comparison holds across controllers rather than in absolute terms.

\subsection{Attention-enhanced LSTM Policy Architecture}
\label{arch}
To address the \textit{information bottleneck} observed in standard RNNs, where fixed-size hidden states struggle to retain high-frequency details within their processed sequence, we implement a custom feature extractor. The raw state vector $\mathbf{s}_t$ is first projected by a two-layer fully connected network (128 units per layer, ReLU) to an embedding dimension $d=128$. This embedding sequence is then processed by a two-layer stacked LSTM with hidden size $d_h=256$ and inter-layer dropout $p=0.1$. To enable the policy to weigh historical states by their relevance rather than averaging them uniformly, we apply a deterministic soft-attention mechanism over the LSTM hidden states $H = \{\mathbf{h}_1, \dots, \mathbf{h}_{w_a}\}$. The attention score $e_i$ for each timestep is computed via a learned linear transformation, given in Eq.~\eqref{eq:6}.
\begin{equation}
    e_i = \mathbf{v}_a^\top \mathbf{h}_i + b_a
    \label{eq:6}
\end{equation}
\begin{equation}
    \alpha_i = \frac{\exp(e_i)}{\sum_{j=1}^{w_a} \exp(e_j)}
    \label{eq:7}
\end{equation}
\begin{equation}
    \mathbf{c}_{ctx} = \sum_{i=1}^{w_a} \alpha_i \mathbf{h}_i
    \label{eq:8}
\end{equation}
\noindent where $\mathbf{v}_a \in \mathbb{R}^{256}$ and $b_a$ are learnable weights. The resulting context vector $\mathbf{c}_{ctx}$, i.e., a weighted sum of the history as given in Eq.~\eqref{eq:8}, is passed through a final projection layer (256 units, ReLU) to form the shared feature embedding consumed by the policy and value heads. This architecture lets the agent react proactively to precursors of instability by weighting the recent window non-uniformly, rather than compressing it into a single fixed-size summary.

The deterministic soft attention operates over a fixed sequence length of $w_a=8$ time steps (eight consecutive 60-second control intervals), and is applied identically during both training and inference, as it is an integral component of the feature extractor forward pass shared by the actor and critic heads. Furthermore, the hidden states $\mathbf{h}_0$ and cell states $\mathbf{c}_0$ of both LSTM layers are re-initialized to zero at every control step. The network is a stateless sequence encoder applied fresh to the current $w_a=8$ observation window, rather than a persistently recurrent policy. No hidden state is carried across control steps or trace-segment boundaries. All temporal context available to the policy comes from the fixed window of raw observations itself. The weights $\boldsymbol{\alpha}$ are not prescribed, they are learned end-to-end with the policy. Section~\ref{sec:attention_analysis} reports the profile recovered at evaluation time on both clusters.

\begin{algorithm}[!t]
\caption{Attention-Enhanced Double-Stacked LSTM-PPO}
\label{alg:attn_ppo_control}
\begin{algorithmic}[1]
\REQUIRE Update interval $\Delta t$, Smoothing window $w_f$, Attention window $w_a$, Discount $\gamma$, Clip $\epsilon$.
\ENSURE Optimized Policy $\pi_\theta$, Value Function $V_\phi$.
\STATE \textbf{Initialize:} Neural networks $\theta, \phi$ and Buffer $\mathcal{B} \leftarrow \emptyset$.
\FOR{episode $k = 1, \dots, K$}
    \STATE Reset environment for episode $k$
    \STATE \textbf{Initialize:} window buffer $\mathcal{W} \leftarrow \emptyset$ \COMMENT{holds the $w_a$ most recent raw states}
    \STATE $t \leftarrow 0$
    \WHILE{$t < T_{max}$}
        \STATE \textbf{State Encoding (Sec.~\ref{obs})}
        \STATE Collect metrics $\mathbf{o}_t = [\mathbf{m}^{perf}, \mathbf{m}^{res}, \mathbf{m}^{conf}]$
        \STATE $\hat{N}_t \leftarrow \text{Smooth}(\lambda_{t-w_f+1:t})$ \COMMENT{Estimate short-horizon demand from last $w_f$ steps}
        \STATE $\mathbf{s}_t \leftarrow [\mathbf{o}_t, \mathbf{x}^{time}_t, \hat{N}_t]$ \COMMENT{Construct full state}
        \STATE Append $\mathbf{s}_t$ to $\mathcal{W}$, dropping the oldest entry if $|\mathcal{W}| > w_a$
        \STATE \textbf{Temporal Processing (Sec.~\ref{arch})}
        \STATE $\mathbf{e}_i \leftarrow \text{ReLU}(\mathbf{W}_e \mathbf{s}_i + \mathbf{b}_e)$ for each $\mathbf{s}_i \in \mathcal{W}$ \COMMENT{Embed every frame in the window}
        \STATE \textbf{Re-initialize} $\mathbf{h}^{(l)}_0, \mathbf{c}^{(l)}_0 \leftarrow \mathbf{0}$ for layers $l \in \{1,2\}$ \COMMENT{Fresh encode; no state carried from $t-1$}
        \STATE $\{\mathbf{h}^{(1)}_i\} \leftarrow \text{LSTM}^{(1)}(\{\mathbf{e}_i\}; \mathbf{h}^{(1)}_0, \mathbf{c}^{(1)}_0)$ \COMMENT{Encode the whole window}
        \STATE $\{\mathbf{h}^{(2)}_i\} \leftarrow \text{LSTM}^{(2)}(\{\mathbf{h}^{(1)}_i\}; \mathbf{h}^{(2)}_0, \mathbf{c}^{(2)}_0)$ \COMMENT{Deep history tracking within the window}
        \STATE \textbf{Attention Mechanism (Sec.~\ref{arch})}
        \STATE Compute attention scores $\mathbf{e}_{attn}$ from $\{\mathbf{h}^{(2)}_i\}$
        \STATE $\boldsymbol{\alpha} \leftarrow \text{softmax}(\mathbf{e}_{attn})$ \COMMENT{Attention weights}
        \STATE $\mathbf{c}_{ctx} \leftarrow \sum_i \alpha_i \, \mathbf{h}^{(2)}_i$ \COMMENT{Context vector}
        \STATE \textbf{Decision Making (Sec.~\ref{action_space})}
        \STATE Sample action $\mathbf{a}_t \sim \pi_\theta(\cdot \mid \mathbf{c}_{ctx})$ \COMMENT{Policy inference}
        \STATE Apply $\mathcal{C}_{k8s}$ to cluster; Wait $\Delta t$
        \STATE \textbf{Feedback Loop (Sec.~\ref{reward})}
        \STATE Measure reward $R_t$ based on SLOs \& Stability
        \STATE Store transition $(\mathbf{s}_t, \mathbf{a}_t, R_t)$ in Buffer $\mathcal{B}$
        \STATE \textbf{PPO Learning Step}
        \IF{$|\mathcal{B}| \ge N_{batch}$}
            \STATE Calculate advantages $\hat{A}_t$ and returns using $V_\phi$
            \FOR{epoch $j=1 \dots M$}
                \STATE $L_{clip} \leftarrow$ PPO clipped surrogate loss
                \STATE $L_{vf} \leftarrow$ MSE$(V_\phi, \text{returns})$
                \STATE Optimize $\theta, \phi$ to minimize $(L_{clip} + L_{vf} - \text{Entropy})$
            \ENDFOR
            \STATE Clear Buffer $\mathcal{B}$
        \ENDIF
        \STATE $t \leftarrow t + 1$
    \ENDWHILE
\ENDFOR
\end{algorithmic}
\end{algorithm}

The complete execution flow of the proposed agent is given in Algorithm~\ref{alg:attn_ppo_control} and illustrated in Fig.~\ref{fig:arch}. This algorithm formalizes the sequential interaction between the environment and the agent, explicitly detailing the forward pass through the double-stacked LSTM layers and the computation of the context vector via the soft-attention mechanism. Furthermore, it also illustrates how the PPO update law is integrated within the training loop to iteratively refine the policy parameters $\theta$ and value function weights $\phi$ based on the collected trajectories in buffer $\mathcal{B}$.

\section{Experimental Results}
\label{results}
\subsection{Hardware and Cluster Testbed}
\label{testbed-description}
Experiments are conducted on two independent Kubernetes clusters hosted within a virtualized lab environment and connected via a low-latency internal network. Every agent is evaluated on both, so that the reported behaviour can be separated from the particular amount of scaling headroom a single testbed happens to provide.

Cluster~1 (C1) is a five-node MicroK8s (v1.31) deployment. Its control node provides 16 vCPUs, 48\,GB RAM, and an NVIDIA L40S GPU hosting the Kubernetes control plane and executes the heavy-duty RL training loops including LSTM/attention inference, offloading complex policy updates to the GPU to prevent CPU overhead. Four worker nodes (4 vCPUs and 16\,GB RAM each) form the execution plane, providing 16 vCPUs of aggregate worker capacity distributed across four independently schedulable hosts.

Cluster~2 (C2) is a two-node deployment (Kubernetes v1.31) whose control node is provisioned identically to C1's (16 vCPUs, 48\,GB RAM, NVIDIA L40S), paired with a single worker node of 16 vCPUs and 64\,GB RAM. C2 therefore offers the same aggregate worker CPU capacity as C1 but concentrated in one host, so a scale-out decision cannot be absorbed by spreading replicas across nodes.

The \texttt{factorizator} pods run under the Kubernetes \emph{Burstable} QoS class (CPU/memory requests of 250\,m/128\,Mi against limits of 500\,m/256\,Mi). Resource telemetry ($u^{cpu}_t$, $u^{ram}_t$) is sampled once per 60\,s control step via \texttt{kubectl top pod}, backed by the cluster's metrics-server.

To ensure high-fidelity workload reproduction, we replay real Azure Functions invocation traces \citep{azurefunctions2021} using the \textit{hey} load generator \citep{hey} within a discrete-time control loop. We randomly sample 7 days from the trace using a fixed random seed (42), and use 5 days for training and 2 days for testing. Each day is discretized into 500 control intervals, yielding 2,500 training and 1,000 testing timesteps with a fixed control interval of 60\,s per step. Moreover, the learning-based methods are also trained and evaluated on five more random seeds, i.e., 123, 456, 789, 1337, and 2024. Since the seed also selects the trace segment, each seed carries a different demand (per-seed train/test load figures are provided alongside the released code~\citep{shaikh_github_2026}). The mean test load spans 4.7$\times$ across seeds, and seed 789 trains on 4.2$\times$ the load it is tested on. Agent-to-agent comparisons are therefore made within a seed, where the demand sequence is identical for every agent.

At each step $t$, the controller computes the target request rate $\bar{N}_t$ from the trace and configures \texttt{hey} to inject the corresponding load into the OpenFaaS \citep{openfaas} gateway, targeting a CPU-bound \texttt{factorizator} serverless function whose deterministic compute profile makes latency changes attributable primarily to scaling decisions. The maximum replica count $R_{max}=100$ (Table~\ref{tab:hyperparameters_grid}) is sized against the heaviest test peak across all six seeds, 2{,}587 req/min, rather than against any single seed. A bound calibrated only to the original, lighter single-seed evaluation would have left the agent without sufficient headroom under the heavier seeds now included. The parameters and hyper-parameters are reported in Table~\ref{tab:hyperparameters_grid}.

\begin{table}[!t]
\caption{Simulation Parameters and Hyperparameters}
\label{tab:hyperparameters_grid}
\centering
\footnotesize
\setlength{\tabcolsep}{3pt}
\renewcommand{\arraystretch}{1.15}
\begin{tabular}{l c l}
\toprule
\textbf{Parameter} & \textbf{Symbol} & \textbf{Value} \\
\midrule
\multicolumn{3}{l}{\textbf{Environment \& Constraints}} \\
\midrule
Control Interval & $\Delta t$ & \SI{60}{\second} \\
Target Latency & $L_{target}$ & \SI{20}{\milli\second} \\
Violation Threshold & $L_{thresh}$ & \SI{50}{\milli\second} \\
Max Replicas & $R_{max}$ & 100 \\
Smoothing Window & $w_f$ & 3 steps \\
Attention/LSTM Window & $w_a$ & 8 steps \\
State Dimension & $dim(\mathcal{S})$ & 13 \\
Action Dimension & $dim(\mathcal{A})$ & 2 (Multi-Discrete) \\
\midrule
\multicolumn{3}{l}{\textbf{Reward Function Weights}} \\
\midrule
SLO Compliance & $w_{slo}$ & 0.55 \\
Resource Efficiency & $w_{cpu}$ & 0.25 \\
Success Ratio & $w_{succ}$ & 0.12 \\
Stability & $w_{stab}$ & 0.08 \\
\midrule
\multicolumn{3}{l}{\textbf{PPO Optimization}} \\
\midrule
Discount Factor & $\gamma$ & 0.99 \\
GAE Parameter & $\lambda$ & 0.93 \\
Entropy Coeff. & $c_{ent}$ & 0.01 \\
Learning Rate & $\alpha$ & \makecell[l]{$2 \times 10^{-4} \to 0$ \\ (cosine decay)} \\
Mini-batch Size & $B$ & 128 \\
Rollout Buffer & $N_{steps}$ & 512 \\
Update Epochs & $K$ & 10 \\
\midrule
\multicolumn{3}{l}{\textbf{Neural Network Architecture}} \\
\midrule
Embedding Dimension & $d$ & 128 (2 FC layers, ReLU) \\
LSTM Hidden Dimension & $d_h$ & 256 \\
LSTM Layers & $N_{stack}$ & 2 (Stacked) \\
Dropout & $p_{drop}$ & 0.1 \\
Attention & - & Soft-Attention \\
Shared Feature Dimension & $d_{feat}$ & 256 (ReLU) \\
Optimizer & - & Adam \\
\bottomrule
\end{tabular}
\end{table}

\subsection{Comparative Experimental Framework}
We conduct a comparative evaluation against three baselines under identical environmental conditions to validate the effectiveness of the proposed Attention-enhanced PPO. For fairness, all \emph{RL-based} controllers share the identical action-decoding and runtime stabilization pipeline (same mapping/clipping rules); only the policy network differs. For the HPA baseline, RL-specific actuators are not used and are held constant throughout (gateway throughput multiplier $=1.0$), while the rest of the testbed and workload replay remain identical. These strategies are selected to isolate the contributions of specific architectural components (ablation study) and to benchmark against alternative RL paradigms:

\begin{figure}
    \centering
    \includegraphics[width=\linewidth]{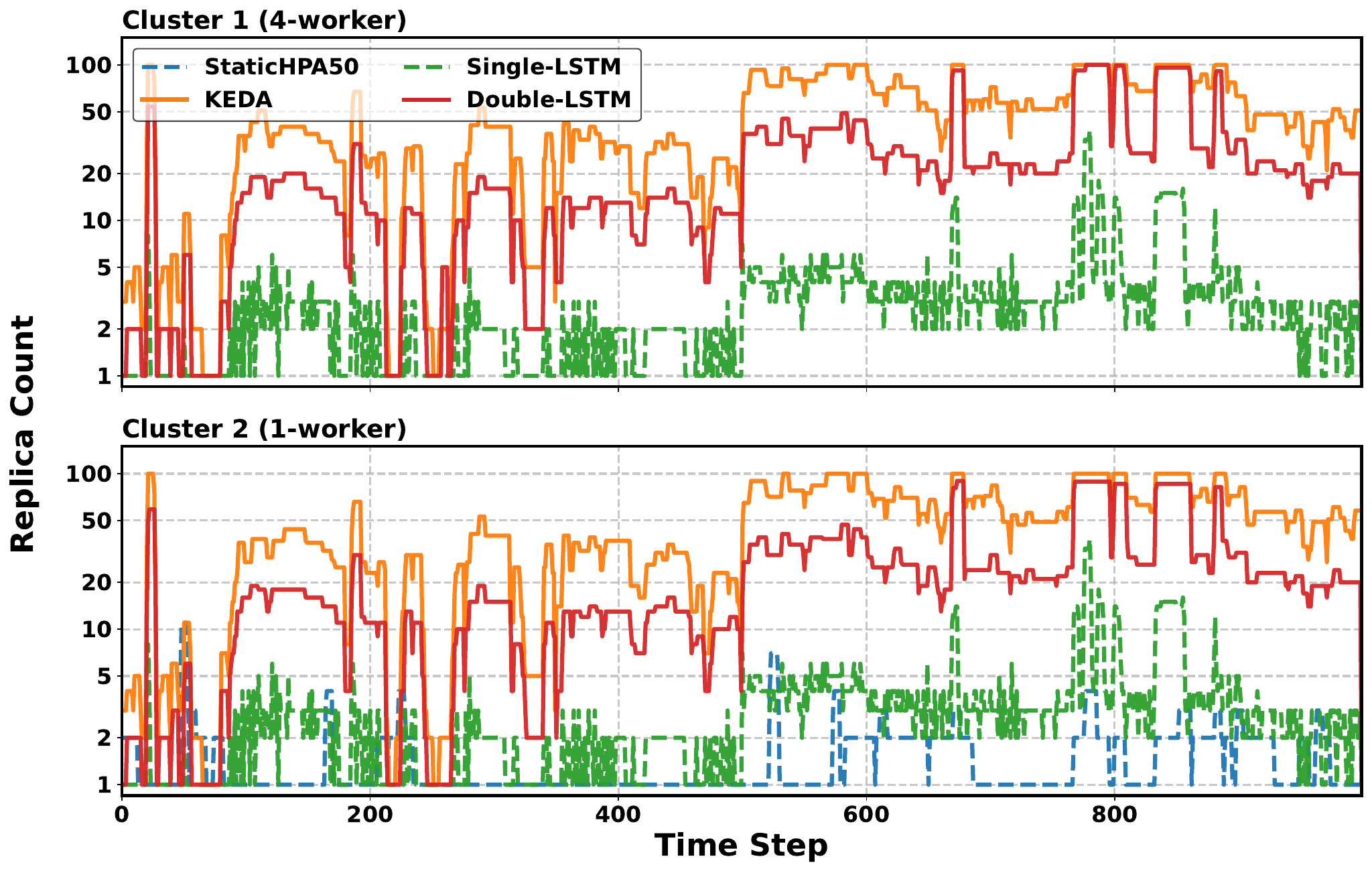}
    \caption{Active replica count, at an identical seed, allocated by each method over the 1{,}000-step evaluation, for Cluster 1 (top) and Cluster 2 (bottom), under a single common seed.}
    \label{fig:replicas}
\end{figure}

\begin{itemize}
    \item Proposed (Attn-LSTM-PPO): The complete architecture is detailed in Section~\ref{arch}. This agent uses a Double-Stacked LSTM for temporal feature extraction and a soft-attention mechanism to identify critical historical precursors, operating on the full 13-dimensional state space with all actuation parameters enabled.
    \item Baseline 1: Recurrent PPO (Ablation Study): An internal ablation of the proposed agent with the soft-attention module disabled, retaining only a single-layer LSTM backbone. This baseline serves as a direct ablation study to isolate the performance gain achieved specifically by the attention mechanism in mitigating the information bottleneck within the attention window.
    \item Baseline 2: KEDA (Reactive, Industry-Standard RL-Free Baseline): KEDA~\citep{keda}, deployed with its HTTP add-on, routes traffic through an interceptor that scales replicas against a fixed target request rate per replica. Unlike the learned baseline, KEDA requires no training and represents the current industry-standard approach to event-driven autoscaling. We evaluate it under load parity with the aux-equipped learned agents (matched throughput multiplier) and calibrate the interceptor's proxy overhead directly against a no-autoscaler baseline to keep the comparison fair.
    \item Baseline 3: Standard HPA (Reactive Benchmark): The industry-standard Kubernetes HPA, configured with a static $50\%$ CPU utilization target, a quite common value, using upstream defaults otherwise: 15\,s sync period, 0.1 tolerance, min/max replicas of 1/100, and a 30\,s scale-down stabilization window. This serves as the control baseline for reactive, PID-style orchestration \citep{10.1145/3465630,10.1145/2890784}.
\end{itemize}

\begin{figure}
    \centering
    \includegraphics[width=\linewidth]{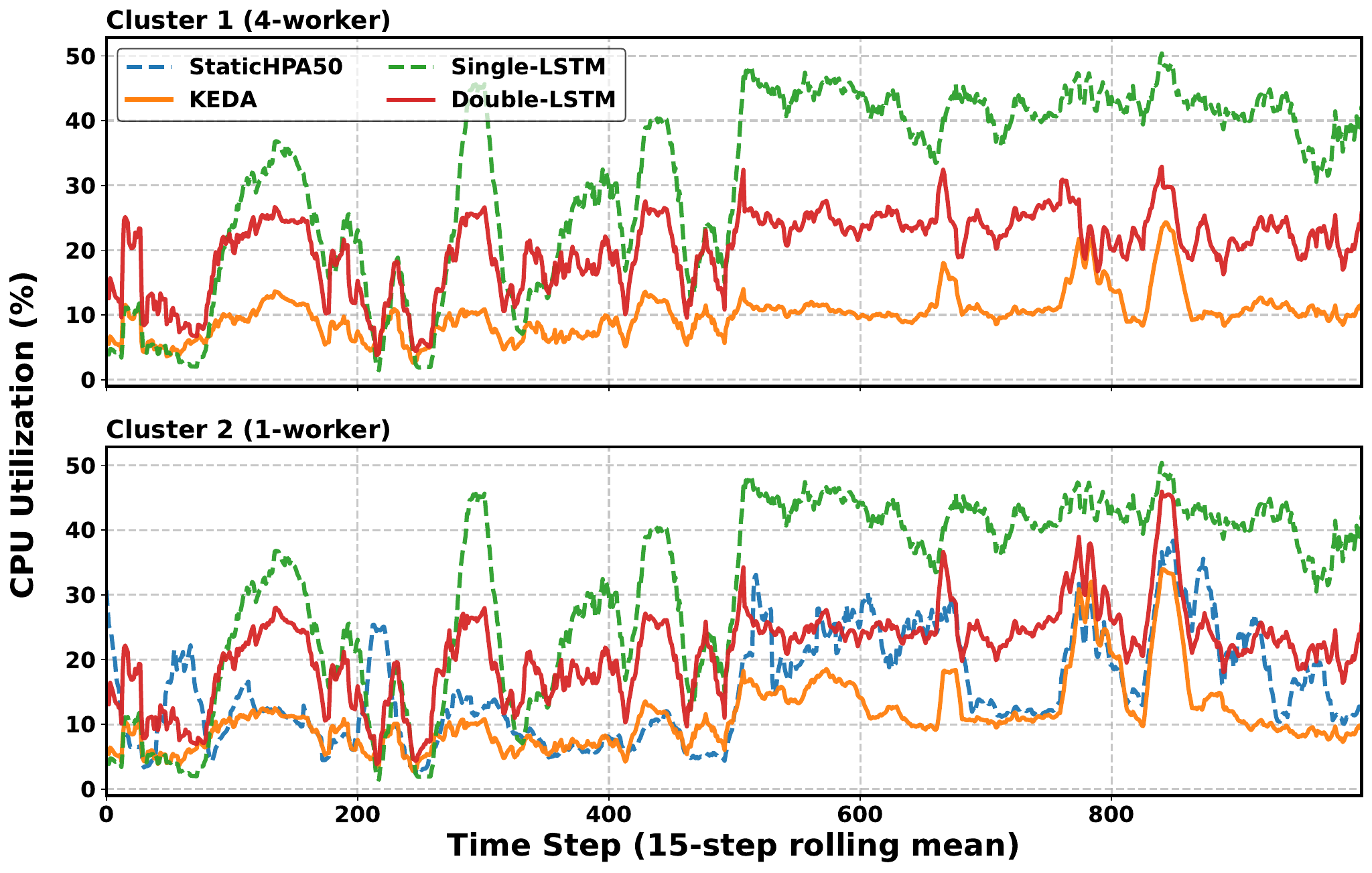}
    \caption{CPU utilization relative to the pod's CPU request (250\,millicores); 200\% corresponds to the container's CPU limit (500\,millicores), for each agent (15-step rolling mean), for Cluster 1 (top) and Cluster 2 (bottom), under a single common seed.}
    \label{fig:cpu}
\end{figure}
%\FloatBarrier
\subsection{Evaluation Protocol and Statistical Treatment}
The agents are not evaluated with the same number of runs. Single-LSTM and Double-LSTM are learning agents for which the final policy depends on the seed, which governs initialization, exploration and the train/test day split (Section~\ref{testbed-description}). Each learning agent is trained and evaluated using five seeds on each cluster, yielding ten seed--cluster evaluation runs. Every aggregate run is reported as a mean with a 95\% confidence interval over those ten runs in Table~\ref{tab:performance_summary} and the following figures. StaticHPA50 and KEDA hold no learned parameters, so repeated executions would resample environmental noise rather than policy variance. KEDA is evaluated once per cluster at seed 42 and reported directly. StaticHPA50 is evaluated once, on C2 only, at seed 42, as the reactive rule-based reference point. Only the Double-LSTM against Single-LSTM comparison is therefore paired, and only it admits a significance test.

Over the $n=10$ matched runs, a paired $t$-test and an exact Wilcoxon signed-rank test separate Double-LSTM from the Single-LSTM ablation on average latency ($p=0.0004$), P90 latency ($p=0.0004$), hard-SLO compliance ($p=0.0001$) and mean reward ($p<0.0001$). For each outcome, the exact two-sided Wilcoxon signed-rank test yields $p=0.0020$, the smallest attainable $p$-value for $n=10$. In every matched seed--cluster pair, Double-LSTM has lower average and P90 latency, higher hard-SLO compliance and reward, and higher replica churn than Single-LSTM. Replica churn separates ($p<0.0001$) in favour of the ablation, which under-provisions and barely moves.

\subsection{Comparative Telemetry Analysis}
To evaluate the operational usage of the proposed architecture, we conducted a comparative telemetry analysis of the Double-LSTM agent against the Single-LSTM ablation, KEDA, and the industry-standard Static HPA. Importantly, all presented figures and performance metrics are derived exclusively from the evaluation (test) phase with exploratory learning disabled, and do not include the initial training phase. Figures~\ref{fig:replicas} through~\ref{fig:latency_cdf} visualize the performance dynamics over a 1,000-step evaluation window on unseen data, isolating the critical trade-offs between provisioning stability, resource efficiency, and service level assurance.

Further, performance metrics for each agent are summarized in Table~\ref{tab:performance_summary} to support reproducibility. This table reports average CPU utilization, mean and tail latency (P50, P90, P95, and P99), reward, SLO compliance at 20\,ms and 50\,ms, average replica count, and cumulative replica churn.

\begin{table*}[htbp]
\centering
\LARGE
\caption{Performance Summary of Evaluated Agents across Resource Utilization, Latency, Reward, SLO Compliance, and Provisioning Stability.}
\label{tab:performance_summary}
\resizebox{\textwidth}{!}{%
\begin{threeparttable}
\setlength{\tabcolsep}{3pt}
\renewcommand{\arraystretch}{1.20}
\begin{tabular}{l c c c c c c c c c c c}
\toprule
\multirow{2}{*}{\textbf{Agent}\tnote{$\ddagger$}} & \multirow{2}{*}{\textbf{Avg CPU (\%)}} & \multicolumn{5}{c}{\textbf{Latency (ms)}} & \multirow{2}{*}{\textbf{Reward}} & \multicolumn{2}{c}{\textbf{SLO Compliance}\tnote{$\S$}} & \multirow{2}{*}{\textbf{Replicas}} & \multirow{2}{*}{\textbf{\shortstack{Replica\\Churn}}} \\
\cmidrule(lr){3-7} \cmidrule(lr){9-10}
 & & \textbf{Avg} & \textbf{P50} & \textbf{P90} & \textbf{P95} & \textbf{P99} & & \textbf{20\,ms} & \textbf{50\,ms} & & \\
\midrule
StaticHPA50 (C2) & 15.44 & 58.82 & 52.20 & 100.2 & ----- & 179.7 & ----- & 10.3\% & 43.5\% & 1.52 $\pm$ 1.09 & 97 \\
Double-LSTM (no-aux) (C2)\tnote{$\dagger$} & 18.80 & 26.99 & 25.61 & 44.5 & 50.16 & 79.8 & 0.637 & 10.8\% & 97.5\% & 8.85 $\pm$ 10.78 & 611 \\
Double-LSTM $w_a$=1 (C1/C2) & 55.78 / 55.43\tnote{$\star$} & 28.36 / 29.13 & 25.42 / 26.09 & 42.58 / 44.04 & 50.60 / 52.71 & 73.77 / 75.83 & 0.59 / 0.57 & 11.1\% / 10.2\% & 97.7\% / 97.8\% & 7.37 $\pm$ 6.29 / 7.27 $\pm$ 5.91 & 409 / 396 \\
\midrule[0.3pt]
KEDA (C1/C2) & 9.95 / 11.16 & 21.05 / 21.62 & 19.16 / 19.33 & 27.3 / 29.5 & 31.16 / 33.71 & 34.7 / 37.6 & 0.635 / 0.620 & 33.2\% / 15.5\% & 100.0\% / 100.0\% & 47.22 $\pm$ 30.49 / 47.77 $\pm$ 30.31 & 2230 / 2325 \\
Single-LSTM (C1) Mean$\pm$95\% CI & 13.98 $\pm$ 9.18 & 57.21 $\pm$ 22.79 & 51.29 $\pm$ 20.09 & 95.3 $\pm$ 44.1 & 108.97 $\pm$ 50.46 & 165.8 $\pm$ 70.4 & 0.331 $\pm$ 0.202 & 25.9\% $\pm$ 20.6\% & 50.2\% $\pm$ 31.7\% & 1.25 $\pm$ 0.71 (CI $\pm$0.13) & 50 $\pm$ 23 \\
Single-LSTM (C2) Mean$\pm$95\% CI & 14.22 $\pm$ 9.32 & 62.52 $\pm$ 27.95 & 56.59 $\pm$ 25.40 & 103.3 $\pm$ 51.3 & 118.07 $\pm$ 58.60 & 183.2 $\pm$ 87.5 & 0.304 $\pm$ 0.215 & 25.8\% $\pm$ 20.6\% & 49.0\% $\pm$ 31.8\% & 1.26 $\pm$ 0.78 (CI $\pm$0.09) & 54 $\pm$ 25 \\
Single-LSTM (C1+C2) Mean$\pm$95\% CI & 14.10 $\pm$ 5.03 & 59.87 $\pm$ 14.00 & 53.94 $\pm$ 12.61 & 99.3 $\pm$ 26.2 & 113.52 $\pm$ 29.90 & 174.5 $\pm$ 43.6 & 0.317 $\pm$ 0.114 & 25.8\% $\pm$ 11.2\% & 49.6\% $\pm$ 17.3\% & 1.26 $\pm$ 0.74 (CI $\pm$0.06) & 52 $\pm$ 13 \\
Double-LSTM (C1) Mean$\pm$95\% CI & 16.19 $\pm$ 5.48 & 21.88 $\pm$ 1.12 & 19.40 $\pm$ 1.52 & 30.6 $\pm$ 0.7 & 34.93 $\pm$ 0.81 & 55.4 $\pm$ 4.1 & \textbf{0.750 $\pm$ 0.018} & 44.5\% $\pm$ 9.2\% & 99.3\% $\pm$ 0.5\% & 12.56 $\pm$ 14.48 (CI $\pm$10.08) & 1031 $\pm$ 386 \\
Double-LSTM (C2) Mean$\pm$95\% CI & 15.46 $\pm$ 6.27 & 24.31 $\pm$ 3.11 & 21.50 $\pm$ 2.92 & 35.3 $\pm$ 5.8 & 40.41 $\pm$ 6.60 & 62.9 $\pm$ 9.8 & 0.714 $\pm$ 0.052 & 32.2\% $\pm$ 18.3\% & 98.3\% $\pm$ 2.2\% & 10.84 $\pm$ 13.15 (CI $\pm$10.89) & 827 $\pm$ 426 \\
Double-LSTM (C1+C2) Mean$\pm$95\% CI & 15.83 $\pm$ 3.21 & 23.10 $\pm$ 1.57 & 20.45 $\pm$ 1.58 & 33.0 $\pm$ 2.9 & 37.67 $\pm$ 3.29 & 59.1 $\pm$ 5.0 & \textbf{0.732 $\pm$ 0.025} & 38.4\% $\pm$ 9.1\% & 98.8\% $\pm$ 0.9\% & 11.70 $\pm$ 13.86 (CI $\pm$5.74) & 929 $\pm$ 234 \\
\bottomrule
\end{tabular}
\begin{tablenotes}
\item[$\dagger$] Same trained agent with throughput multiplier fixed at $1\times$.
\item[$\star$] With one frame of context, the Double-LSTM agent uses a 90\% CPU-utilization target on both C1 and C2. Relative to the full-context configuration, it provisions approximately one third fewer replicas, concentrating the admitted load on fewer pods and yielding average CPU utilization of approximately 55\%.
\item[$\ddagger$] For Single-LSTM and Double-LSTM, C1, C2, and C1+C2 entries are means over five, five, and ten seed--cluster runs, 95\% CI. Replicas column reports mean $\pm$ standard deviation of the per-step replica count within each run, with the resulting across-run 95\% CI given separately in parentheses. StaticHPA50, KEDA, and the no-aux/$w_a$=1 rows are each a single deterministic run; any $\pm$ shown for these rows is within-run standard deviation, not a confidence interval.
\item[$\S$] Target- and hard-SLO compliance are computed from per-step mean latency relative to the 20\,ms and 50\,ms thresholds.
\end{tablenotes}
\end{threeparttable}%
}
\end{table*}
\subsubsection{Provisioning Dynamics and Control Stability}
The replica count metric in Fig.~\ref{fig:replicas} reveals fundamental differences in provisioning dynamics. To quantitatively assess this, we analyze the average replica footprint alongside operational jitter. We define replica churn as $\sum_{t=2}^{T} |\rho_t - \rho_{t-1}|$ over the evaluation horizon, where $\rho_t$ is the active replica count at step $t$. Lower churn indicates fewer sudden scale transitions and reduced control oscillation.

As detailed in Table~\ref{tab:performance_summary}, the two reactive baselines sit at opposite ends of the churn range. KEDA scales almost continuously in response to raw request-rate thresholds, repeatedly saturating the 100-replica ceiling and then retreating (Fig.~\ref{fig:replicas}). This behaviour produces the highest churn of any evaluated autoscaler (2230 and 2325 on the two clusters) and the largest average footprint (47.22 $\pm$ 30.49 replicas and 47.77 $\pm$ 30.31), without matching the reward achieved by the proposed Double-LSTM agent. Conversely, the static HPA baseline demonstrates rigid, delayed scaling (churn of 97) that misses macro-level spikes entirely due to its mandatory cooldown hysteresis, holding only 1.52$\pm$1.09 replicas on average.

Between these extremes, the Single-LSTM ablation barely moves at all, holding 1.26$\pm$0.74 replicas on average with a churn of only 52$\pm$13. This is not the oscillatory \textit{thrashing} behavior. It instead reflects the opposite failure mode, chronic under-provisioning. The weakness appears in its high tail latency and low hard-SLO compliance (49.6$\pm$17.3\%; Table~\ref{tab:performance_summary}). In contrast, the proposed Double-LSTM agent moves substantially more, averaging 11.70$\pm$13.86 replicas with a churn of 929$\pm$234. This higher scaling activity is deliberate, enabling the agent to add capacity promptly during the peak demand bursts shown in Fig.~\ref{fig:replicas}. This response occurs with a virtually static HPA setpoint: across all five seeds on both clusters, the policy selects the 30\% target for most of the control steps. Traffic spikes are therefore absorbed by Kubernetes scaling aggressively against a low, persistent threshold, rather than through continuous target modulation by the agent. This policy convergence proves consistent across all random seeds and physical testbeds. In contrast, the Single-LSTM baseline exhibits seed-dependent instability, shifting among 30\%, 50\%, and 70\% setpoints across runs without settling on a coherent operating strategy. By locking in this low target, the controller sustains cluster capacity across the entire burst window, preventing premature scale-down actions during transient traffic dips. This behavior is consistent with its lower mean and tail latency relative to Single-LSTM. These results show that churn alone is not a reliable indicator of stability. Its value depends on whether the underlying scaling decisions are correct.

\subsubsection{Resource Utilization and Efficiency}
\begin{figure}
    \centering
    \includegraphics[width=\linewidth]{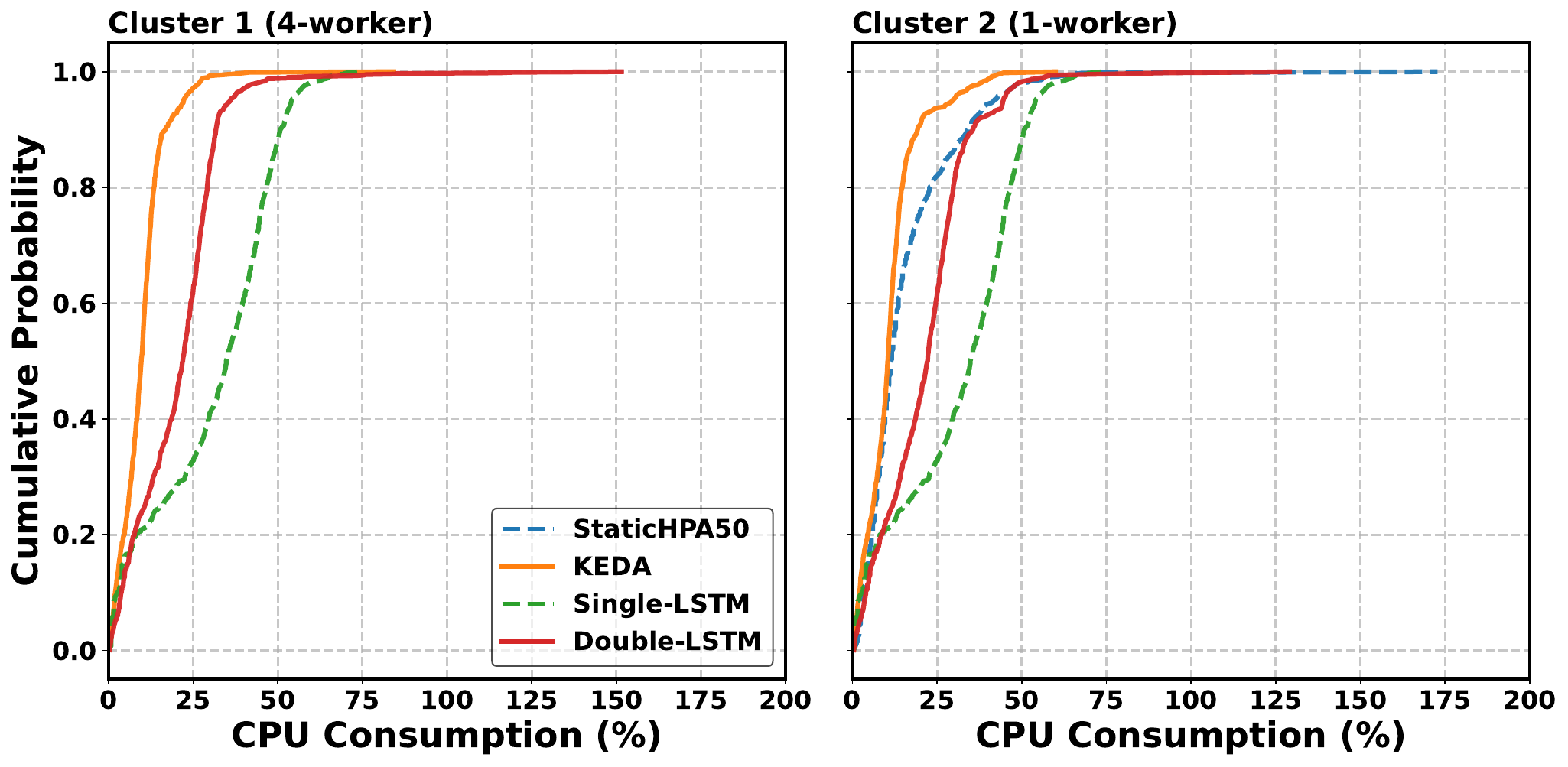}
    \caption{Cumulative Distribution Function (CDF) of per-step CPU utilization (relative to the pod's CPU request, capped at the 200\% CPU-limit ceiling), under a single common seed, for Cluster 1 (left) and Cluster 2 (right). A curve that rises early indicates load spread thinly across many replicas.}
    \label{fig:cpu-cdf}
\end{figure}

Resource efficiency is analyzed through the temporal CPU utilization trends in Fig.~\ref{fig:cpu} and the Cumulative Distribution Function (CDF) in Fig.~\ref{fig:cpu-cdf}. An effective autoscaler seeks high resource usage subject to acceptable latency and SLO compliance, operating at higher safe CPU utilization levels without triggering service degradation. As detailed in Table~\ref{tab:performance_summary}, the two LSTM-based agents record almost similar average utilization once evaluated across five seeds and both testbeds (Double-LSTM 15.83\%$\pm$3.21\%, Single-LSTM 14.10\%$\pm$5.03\%). The replica counts explain why this similarity is misleading. Single-LSTM sustains its utilization on only 1.26$\pm$0.74 replicas, so the offered load concentrates on very few pods and per-pod utilization stays high by default. The agent still fails to convert that utilization into acceptable average latency (59.87$\pm$14.00\,ms), which is the signature of under-provisioning rather than efficient resource use. The Double-LSTM agent reaches a comparable utilization level while running 11.70$\pm$13.86 replicas, and it delivers 23.10$\pm$1.57\,ms average latency at the same time. Although it briefly spikes to very high CPU usage during sudden traffic surges, the system is simply working at full capacity to process pending requests while waiting for new replicas to start up, rather than suffering from continuous overload. As a result, CPU usage returns to its baseline once the additional capacity comes online (Fig.~\ref{fig:cpu}).

%\FloatBarrier

\begin{figure}
    \centering
    \includegraphics[width=\linewidth]{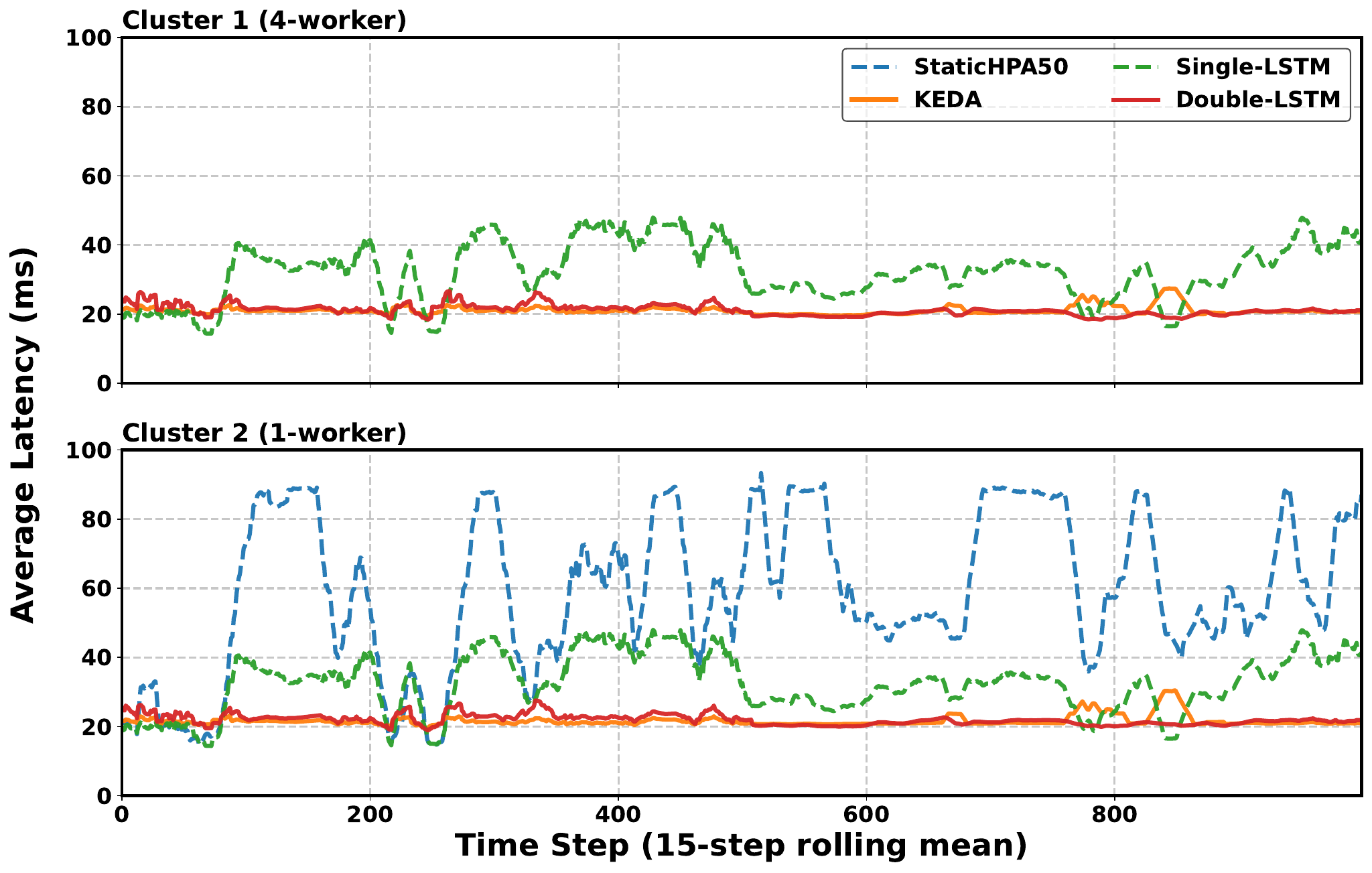}
    \caption{Average request latency (ms) at each evaluation step, under a single common seed, for Cluster 1 (top) and Cluster 2 (bottom), smoothed with a 15-step centered rolling mean.}
    \label{fig:avg_latency}
\end{figure}
%\FloatBarrier
On the other hand, the non-predictive baselines demonstrate resource mismanagement in opposite directions. KEDA records the lowest average utilization of any evaluated autoscaling method (9.95\% on C1 and 11.16\% on C2) and never exceeds 100\% CPU at any evaluation step (Fig.~\ref{fig:cpu-cdf}). KEDA maintains 47.22$\pm$30.49 replicas on C1 and 47.77$\pm$30.31 on C2. Its large footprint spreads the offered load across many pods, trading resource efficiency for low utilization and low mean latency. This behaviour is the resource-cost counterpart to its competitive raw latency. The Static HPA baseline shows the opposite pattern. Its average utilization of 15.44\% conceals a bimodal profile, since the autoscaler remains idle at a single or two replicas for most of the horizon and then saturates beyond 100\% during the sharpest spikes, peaking at 172.8\%. The mandatory cooldown windows observed in reactive scalers delay each scale-up, hence the few active containers absorb the entire spike and latency degrades to 58.82\,ms with only 10.3\% target SLO compliance. However, simply decreasing these mandatory cooldown windows in Static HPA, without an additional intelligent control algorithm, is often counterproductive, as it directly leads to increased control instability and resource thrashing \citep{BENEDETTI2026108112}.

\subsubsection{Latency Distribution and Reliability}
Fig.~\ref{fig:avg_latency} visualizes the temporal latency trajectories, whereas Table~\ref{tab:performance_summary} reports their aggregate distributions through mean, P50, P90, P95, and P99 latency. The reactive baselines exhibit different latency profiles. StaticHPA50 (C2) records a mean latency of 58.82\,ms and a P50 latency of 52.20\,ms. Although its P50 latency is already well above the 20\,ms target, its substantially larger P90 and P99 values indicate a striking upper tail during demand ramps (Section~\ref{sec:tail-latency}). In contrast, KEDA achieves lower latency on both clusters: mean latency is 21.05/21.62\,ms and P50 is 19.16/19.33\,ms for C1/C2. This lower-latency distribution coincides with the large replica footprint and high churn reported in Table~\ref{tab:performance_summary}.

Among the learning-based methods, the Single-LSTM ablation has the weakest aggregate latency profile. Across C1 and C2, its mean latency is 59.87$\pm$14.00\,ms, and its P50 latency is 53.94$\pm$12.61\,ms. The elevated median and broad upper tail indicate persistent latency degradation rather than isolated outliers, consistent with the low replica footprint shown in Fig.~\ref{fig:replicas}. The proposed Double-LSTM provides the strongest latency results among the learning-based configurations. Its mean latency is 23.10$\pm$1.57\,ms, and its P50 latency is 20.45$\pm$1.58\,ms. This reduction in both central and upper-tail latency is consistent with the additional recurrent depth and attention weighting described in Section~\ref{arch}, which let the agent provision decisively rather than default to the chronic under-provisioning that limits Single-LSTM (Fig.~\ref{fig:replicas}). The substantially lower variation across the evaluated C1/C2 runs further indicates this behavior holds consistently across both testbeds rather than depending on the particular scaling headroom of one cluster.

The one-frame Double-LSTM configuration ($w_a = 1$) has mean latency of 28.36/29.13\,ms on C1/C2, with P50 values of 25.42/26.09\,ms, P90 values of 42.58/44.04\,ms, P95 values of 50.60/52.71\,ms, and P99 values of 73.77/75.83\,ms. Relative to the full-context agent on the same clusters (21.88/24.31\,ms mean latency, 30.6/35.3\,ms P90, Table~\ref{tab:performance_summary}), collapsing the window to a single frame raises mean latency by 30\%/20\% and P90 latency by 39\%/25\% on C1/C2, while hard-SLO compliance falls from 99.3\%/98.3\% to 97.7\%/97.8\%. Its higher tail latency relative to the full-context Double-LSTM is consistent with its lower provisioning footprint and reduced replica churn, demonstrating the trade-off between conservative provisioning and tail-latency protection.

\begin{figure}
    \centering
    \includegraphics[width=\linewidth]{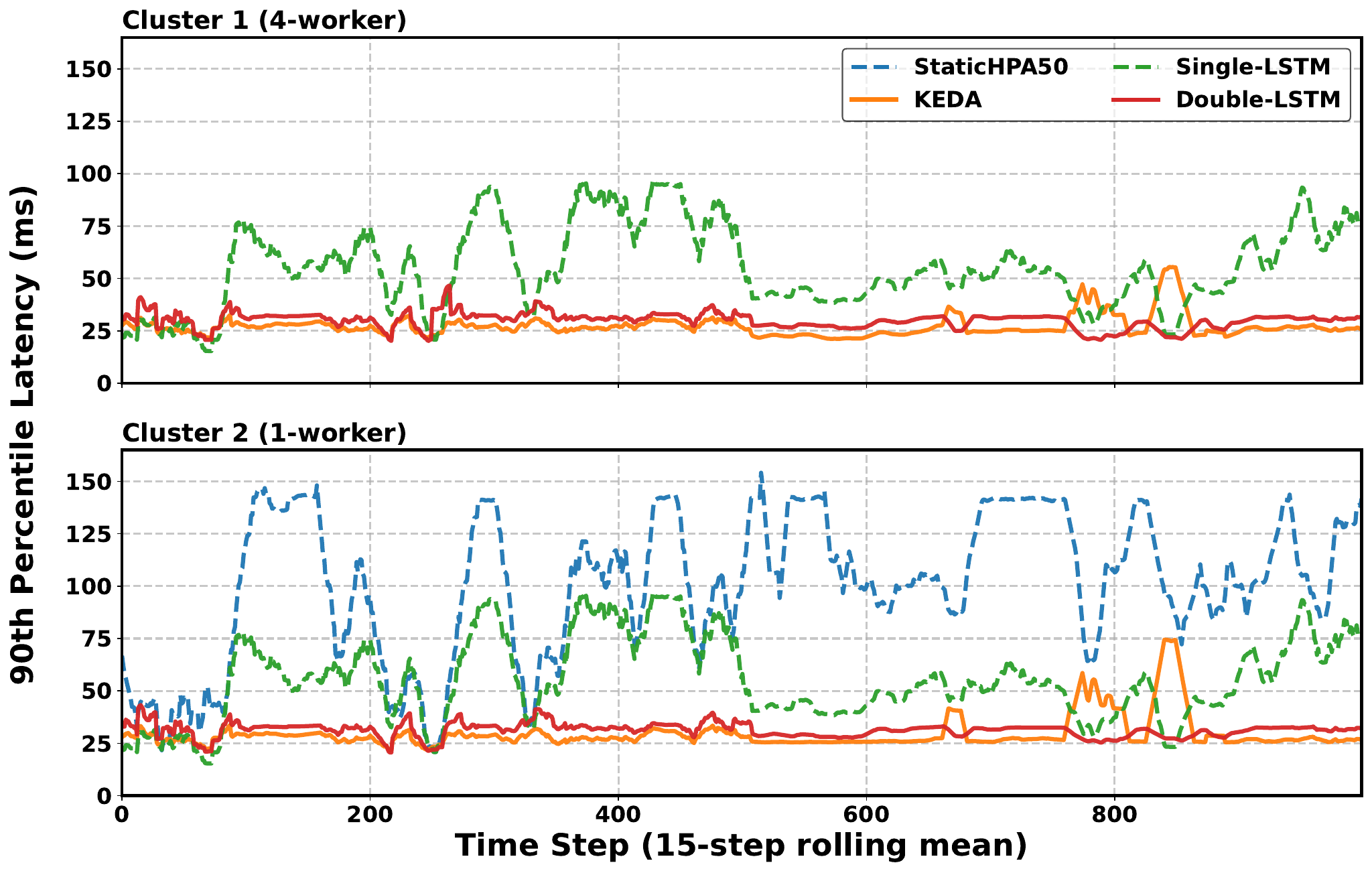}
    \caption{90th percentile (P90) latency, representing the tail performance experienced by the majority of requests, for Cluster 1 (top) and Cluster 2 (bottom), under a single common seed, smoothed with a 15-step centered rolling mean.}
    \label{fig:p90_latency}
\end{figure}
%\FloatBarrier
\subsubsection{Tail Latency Analysis}
\label{sec:tail-latency}
To quantify reliability under stochastic demand, we analyze the 90th percentile (P90) latency. P90 is the latency threshold below which 90\% of requests complete. It is therefore less dominated by the most extreme observations than P99 or maximum latency, while still characterizing the upper-tail experience of a substantial fraction of requests. Figure~\ref{fig:p90_latency} further visualizes the evolution of per-step P90 latency over the evaluation horizon and highlights the limitations of the non-predictive baselines. The Static HPA struggles to adapt to sudden load increases, with the temporal P90 trace exceeding 300\,ms during the primary workload surge near step 520. Its aggregate request-level P90 is 100.2\,ms and its P99 is 179.7\,ms (Table~\ref{tab:performance_summary}), indicating a broad latency tail. This behaviour is consistent with the provisioning delay of reactive threshold-based policies, i.e., capacity becomes available only after load has already increased, leaving incoming requests queued while new replicas are created \citep{xu2025auto}.

KEDA avoids the large sustained P90 excursions observed for StaticHPA50 and records request-level P90 latency of 27.3\,ms on C1 and 29.5\,ms on C2. Its request-level P95 and P99 latencies are 31.16/33.71\,ms and 34.7/37.6\,ms. The temporal P90 trace nevertheless exhibits larger short-lived trips. The 95th percentile \emph{across per-step P90 values} reaches 50.0\,ms on C1 and 62.6\,ms on C2. This latency profile coincides with a large provisioning footprint of 47.22$\pm$30.49 and 47.77$\pm$30.31 replicas on C1 and C2, rather than with predictive demand anticipation.

Among the learning-based methods, the Single-LSTM ablation has the weakest tail-latency distribution across the C1/C2 evaluation runs. The aggregate request-level P90, P95, and P99 are 99.3$\pm$26.2\,ms, 113.52$\pm$29.90\,ms, and 174.5$\pm$43.6\,ms. Further, its per-step P90 remains $\leq 50$\,ms hard-SLO threshold for only 41.2\%$\pm$16.6\% of the evaluation horizon.

The proposed Double-LSTM substantially improves tail latency relative to the Single-LSTM ablation. Its aggregate request-level P90, P95, and P99 are 33.0$\pm$2.9\,ms, 37.67$\pm$3.29\,ms, and 59.1$\pm$5.0\,ms. Moreover, the P95 \emph{across its per-step P90 values} is 38.0\,ms, and its P90 remains at or below the 50\,ms hard-SLO threshold for 95.9\%$\pm$2.8\% of the evaluation window. Thus, the temporal trace indicates that the remaining P90 excursions are short-lived rather than persistent, visible on the considerably narrower aggregate tail-latency distribution relative to Single-LSTM.

Table~\ref{tab:performance_summary} measures SLO compliance using per-step mean latency, under which KEDA never exceeds the 50\,ms hard-SLO bound and Double-LSTM achieves 99.3\%/98.3\% (C1/C2). Here, we report a stricter, tail-focused metric instead: Tail SLO-compliance is the fraction of steps where per-step P90 latency, rather than mean latency, remains at or below the 50\,ms hard SLO, so its values are lower than and not directly comparable to Table~\ref{tab:performance_summary}'s mean-based figures. Under this P90 criterion, the trace in Fig.~\ref{fig:p90_latency} reveals violations on 4.9\% of C1 steps and 5.6\% of C2 steps for KEDA, including bursts of up to 10 and 24 consecutive steps, corresponding to a tail-SLO compliance of 95.1\% and 94.4\%. Double-LSTM achieves comparable or higher tail-SLO compliance under this same criterion (97.6$\pm$2.2\% on C1 and 94.3$\pm$6.3\% on C2), with shorter maximum violation bursts (3/5 versus 10/24 steps) and substantially fewer replicas (12.56/10.84 versus 47.22/47.77). Thus, while KEDA retains a lower aggregate request-level latency distribution, its mean-latency SLO compliance masks the short bursts documented here. Double-LSTM reaches comparable tail-SLO compliance with a substantially smaller replica footprint, a more resource-efficient reliability trade-off.

\subsection{Impact of Recurrent Depth and Attention on Latency Stability}
\label{sec: attention-latency}
\begin{figure}
    \centering
    \includegraphics[width=\linewidth]{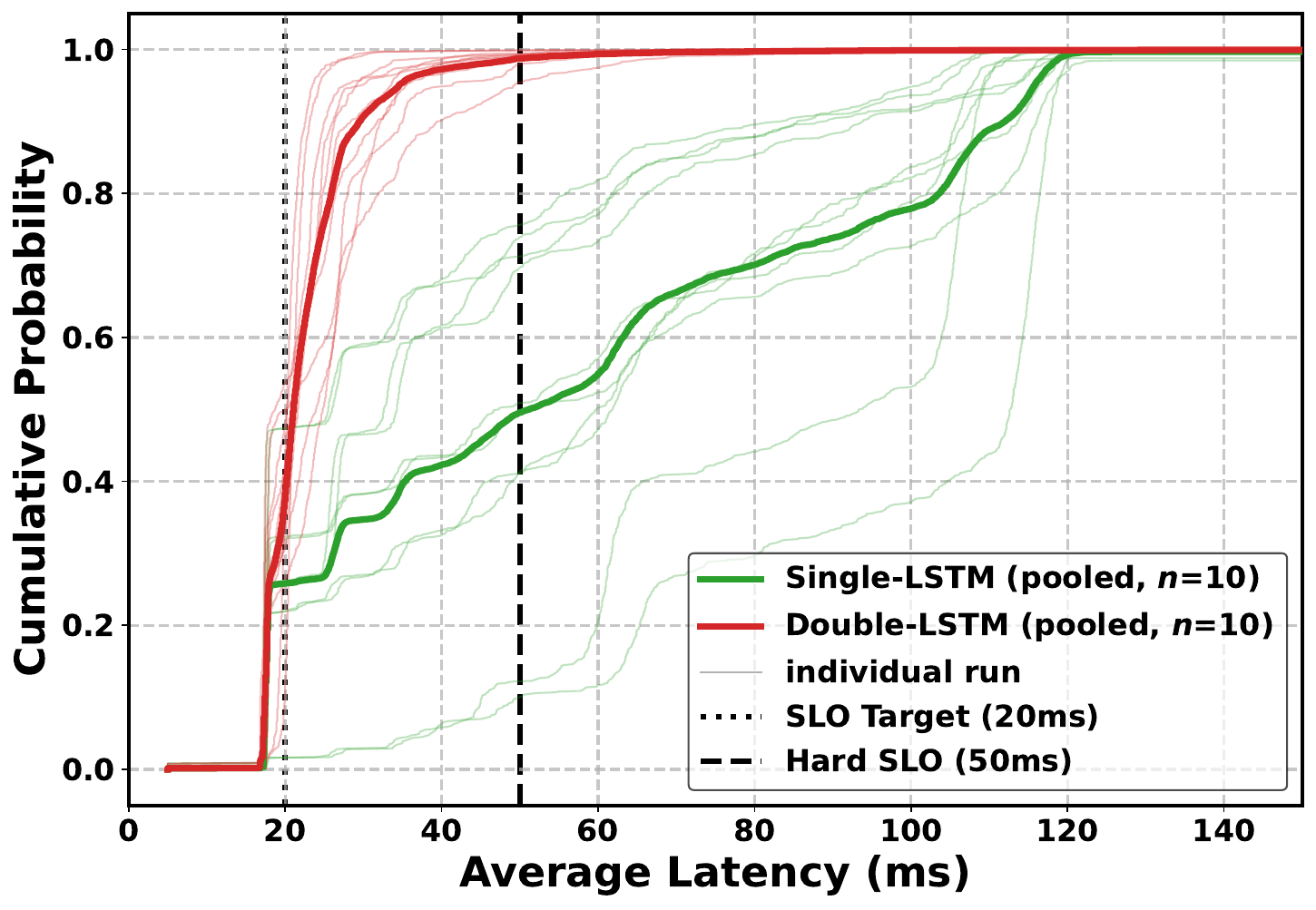}
    \caption{Cumulative distribution function of average request latency over ten matched runs per agent, covering five seeds on each of the two clusters.}
    \label{fig:latency_cdf}
\end{figure}

To isolate the contributions of the proposed architecture, we have conducted an ablation study comparing the Double-LSTM agent against the baseline Single-LSTM variant, as shown in Figure~\ref{fig:latency_cdf}. The two differ in recurrent depth and in the soft-attention module, so the comparison bounds their combined contribution rather than attributing the gain to either factor alone. Both agents are evaluated over the same ten runs, covering five seeds on each of the two clusters, and each seed fixes an identical train and test day split for the two architectures. The two architectures differ first in how much capacity they provision. The Single-LSTM baseline averages 1.26 $\pm$ 0.74 replicas against 11.70 $\pm$ 13.86 for the Double-LSTM agent (Table~\ref{tab:performance_summary}), remaining at two replicas or fewer for 95.9\% of the evaluation horizon against 30.1\% of the horizon for the Double-LSTM agent. The operational consequence of that difference appears in the latency distribution. Here we report interval latency, i.e., the per-control-interval average request latency underlying Fig.~\ref{fig:latency_cdf}, which differs from the per-request percentiles reported in Table~\ref{tab:performance_summary}. The Double-LSTM agent delivers a much tighter profile, with its CDF rising steeply near 20\,ms and reaching 98.8\% compliance at the 50\,ms hard SLO. Its 95th percentile interval latency is 34.5\,ms. Conversely, the Single-LSTM agent shows a long tail, where chronic under-provisioning delays the response to workload spikes. Its mean interval latency of 59.87\,ms already exceeds the hard SLO, it satisfies that threshold in only 49.6\% of steps, and its 95th percentile interval latency reaches 115.8\,ms. The separation holds in every individual run: the poorest Double-LSTM run has a median of 24.8\,ms against 25.5\,ms for the strongest Single-LSTM run, and the Double-LSTM agent attains the lower median in all ten matched pairs (exact sign test, $p < 0.002$).

%\FloatBarrier
\subsection{What the Attention Mechanism Learns}
\label{sec:attention_analysis}
\begin{figure*}[h]
    \centering
    \includegraphics[width=\linewidth]{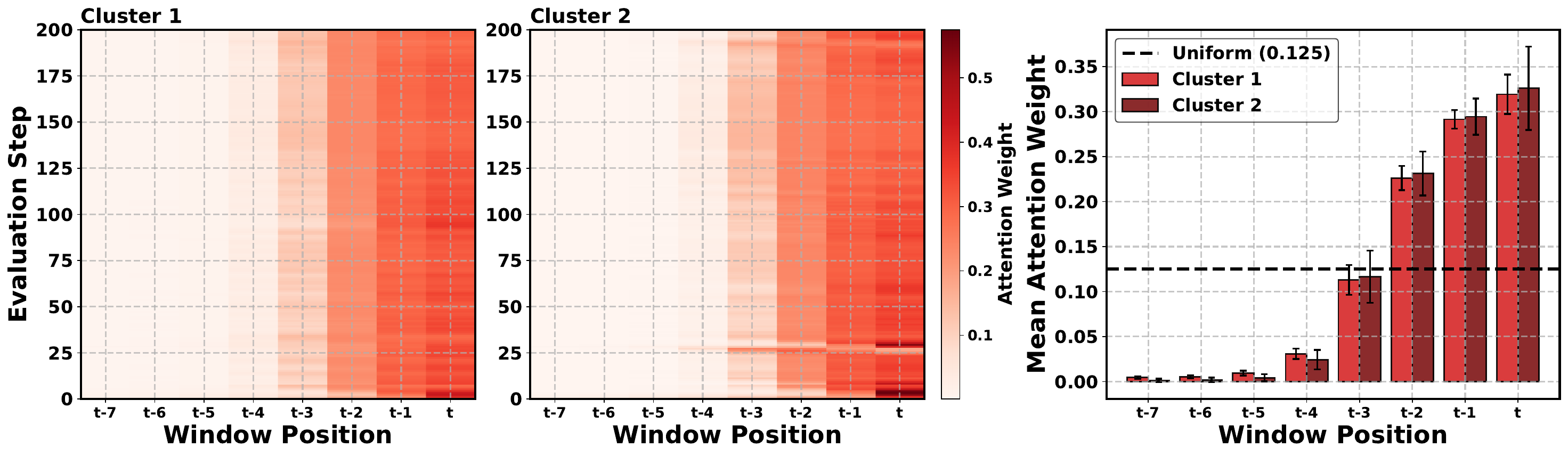}
    \caption{Learned attention weights $\boldsymbol{\alpha}$ over the $w_a=8$ step observation window at evaluation time (first 200 steps shown). Left and centre: per-step weights on Cluster~1 and Cluster~2. Right: mean weight per position with standard deviation, against the uniform reference $1/w_a = 0.125$.}
    \label{fig:attention}
\end{figure*}

Since the attention weights are produced by the same forward pass that selects actions, they can be recorded at every evaluation step. Fig.~\ref{fig:attention} shows a strongly non-uniform, monotonically recency-peaked profile on both clusters. The current interval receives an attention weight of $0.319 \pm 0.022$ (C1) and $0.326 \pm 0.046$ (C2) against the uniform reference of $0.125$, while the oldest position $t-7$ falls to $0.0045$ and $0.0013$. However, it does not collapse onto the most recent frame. The normalized entropy of $0.726$ and $0.697$ corresponds to an effective support of roughly $4.5$ and $4.3$ of the eight available positions, and approximately $68\%$ of the attention mass falls on past frames rather than on the current observation, which is precisely the information a controller acting on a single current observation would discard. The profile is recovered independently on two clusters, reflecting the temporal structure of the workload rather than a single deployment.

The same measurement bounds the claim. Attention places $0.837$ (C1) and $0.852$ (C2) of its mass on the three most recent intervals and never exceeds the uniform reference beyond $t-4$, hence it acts as a learned low-pass filter over the immediate past rather than a mechanism for long-range recall. The gains reported above are consistent with this separation of short-horizon trend from single-step jitter, though Section~\ref{sec: attention-latency} bounds only the combined contribution of depth and attention rather than isolating this mechanism specifically. The unused tail is inexpensive, since the parameter count of the feature extractor depends only on the frame dimension and not on $w_a$, leaving the agent to select its own effective attention span within the provided budget.

%\subsubsection{Window-Length Sensitivity} \label{sec:window_sensitivity}
%The attention profile above motivates a direct check: how sensitive is performance to the attention/LSTM window length $w_a$ itself? 
Further, we evaluated the sensitivity of the trained $w_a=8$ policy by transferring its weights across three alternative window lengths ($w_a \in \{1, 3, 10\}$). Each configuration was tested on both clusters using a single deterministic execution at seed 42, serving as a diagnostic probe rather than a multi-seed benchmark. Table~\ref{tab:window_sensitivity} reports these comparisons, where the $w_a=1$ row reproduces the ablation examined in Section~\ref{sec:tail-latency}. Performance remains practically identical between $w_a=8$ and $w_a=10$ on both clusters, keeping mean latency within 1\,ms and reward within 0.01. This level matches the attention profiles in Fig.~\ref{fig:attention}, where the first four positions capture over 98\% of the cumulative attention mass; once the window encompasses this active horizon, expanding the sequence length provides marginal utility. Conversely, reducing the context to $w_a=1$ degrades all metrics on both testbeds. The intermediate length $w_a=3$ recovers substantial performance over $w_a=1$ on C1, yet yields a slight reward penalty on C2 (0.531 versus 0.565) despite sustaining comparable latency.
\begin{table}[!t]
\caption{Window-length sensitivity (seed 42, single run per cluster; the $w_a=8$ row is the main model from Table~\ref{tab:performance_summary}).}
\label{tab:window_sensitivity}
\centering
\footnotesize
\setlength{\tabcolsep}{3pt}
\resizebox{\columnwidth}{!}{%
\begin{tabular}{c c c c c c}
\toprule
\multirow{2}{*}{$w_a$} & \textbf{Avg Lat (ms)} & \textbf{P90 (ms)} & \textbf{Reward} & \textbf{Replicas} & \textbf{Churn} \\
 & C1 / C2 & C1 / C2 & C1 / C2 & C1 / C2 & C1 / C2 \\
\midrule
1 & 28.36 / 29.13 & 42.58 / 44.04 & 0.59 / 0.57 & 7.37 / 7.27 & 409 / 396 \\
3 & 29.24 / 28.49 & 44.20 / 42.65 & 0.62 / 0.53 & 11.35 / 13.01 & 983 / 1467 \\
8 (main) & 21.16 / 21.96 & 29.97 / 31.31 & 0.75 / 0.74 & 24.51 / 23.63 & 1440 / 1368 \\
10 & 21.07 / 21.90 & 29.61 / 31.11 & 0.75 / 0.74 & 24.58 / 23.76 & 1487 / 1371 \\
\bottomrule
\end{tabular}%
}
\end{table}

\section{Discussion and Lessons Learned}
\label{sec:discussion}
The experimental results presented in Section~\ref{results} support the central premise of this study that mitigating temporal blindness improves control stability under bursty workloads. By unifying short-term demand estimation with control via an Attention-Enhanced Double-Stacked LSTM architecture, our framework delivers consistently lower and tighter latency than both the recurrent ablation and the reactive baselines, and it does so at a fraction of the replica footprint required by the reactive scaler that comes closest to matching it.

The comparative ablation study (Fig.~\ref{fig:latency_cdf}) highlights that memory depth alone is insufficient for robust control. The Single-LSTM baseline performs no better than the static HPA it was intended to improve upon, recording 59.87$\pm$14.00\ ms average latency against the baseline's 58.82\,ms. Its weakness is not merely oscillation. The agent instead settles at two replicas or fewer for 95.9\% of the evaluation horizon, so a single recurrent layer proves able to follow the demand signal yet unable to act on it decisively. Adding a second recurrent layer and a soft-attention mechanism changes the provisioning behaviour itself, and this converts the same temporal information into a 67\% reduction in tail latency (Table~\ref{tab:performance_summary}). This is consistent with reports that weighting historical hidden states helps separate transient variance from genuine trend \citep{meng2023deepscaler}, and with recent findings that attention mechanisms provide a useful inductive bias for structuring temporal dependencies in edge workloads more broadly \citep{zhao2025mscnet}.

A recurring challenge in autoscaling is the zero-sum game between low latency and low oscillation. Reactive controllers, such as the Kubernetes HPA, prioritize stability through hysteresis (already defined in the introduction), resulting in the \textit{reaction latency} observed in Fig.~\ref{fig:p90_latency}. Conversely, threshold-driven scalers that react aggressively to instantaneous load produce the opposite failure, which KEDA exhibits as near-continuous oscillation between low replica counts and the environment's scaling ceiling (Fig.~\ref{fig:replicas}). Our proposed reward function (Section~\ref{reward}), specifically the penalty term $w_{stab} R_{Stab}$ combined with the smoothed policy updates of PPO, allowed the agent to navigate this trade-off effectively. The agent does not scale ahead of demand, since its replica trajectory still follows the workload rather than leading it. What it learns instead is to size each response correctly and to hold that capacity for the duration of a burst rather than retreating during momentary dips, which avoids the flapping characteristic of purely reactive systems \citep{k8s_hpa_flapping}.

While the Double-Stacked LSTM architecture offers better control, it introduces non-negligible computational complexity compared to lightweight heuristics. The inference time for the attention-based network is higher than the simple arithmetic threshold check of standard HPA. In our testbed, this overhead measured 0.70\,ms on CPU and 0.80\,ms on GPU per control step, roughly 0.001\% of the 60\,s control interval and therefore negligible regardless of hardware acceleration. However, in highly constrained edge nodes (e.g., IoT gateways or micro-MECs) without hardware acceleration, the inference latency of deep recurrent networks could potentially compete with the application workload itself. Future deployments may require model quantization or knowledge distillation techniques to reduce the footprint of the policy network for deployment on embedded edge devices \citep{gan2022adaptive}.

It is also important to acknowledge the boundaries of our experimental design. First, while the Azure Functions traces ensure realistic arrival patterns, the use of the Hey load generator in a virtualized environment may not fully capture the \textit{noisy neighbor} interference and hardware contention present in multi-tenant bare-metal clusters. Network I/O contention and CPU cache thrashing, common in production 6G nodes \citep{meybodi2022tedge}, were modeled implicitly using the stochasticity of the environment, however, they were not explicitly controlled variables. Second, the agent was trained and evaluated on a single microservice type (CPU-bound factorization). In real-world microservice chains, an autoscaling decision in one tier (e.g., frontend) can cause back-pressure or starvation in downstream dependencies (e.g., database) \citep{santos2025gwydion}. Our current single-agent formulation does not account for these cascading inter-service dependencies.

Further, a key limitation of our end-to-end architecture is the trade-off between decision transparency and stability. Since the LSTM feature extraction and PPO policy are tightly coupled, diagnosing the root cause of an SLO violation, whether an estimation error or a suboptimal policy action, is highly challenging. Full transparency in DRL remains an open problem, which underscores the importance of a deterministic, bounded translation from policy outputs to Kubernetes configuration values, such as the clipping applied in Section~\ref{action_space}, for mission-critical deployments.

\section{Conclusion}
\label{sec:conclusion}
This article presented a stability-aware autoscaling framework designed to mitigate the \textit{temporal blindness} in standard Reinforcement Learning agents operating within bursty edge environments. By integrating an Attention-enhanced Double-Stacked LSTM architecture into a PPO control loop, our approach sizes each scaling decision against recent history rather than against instantaneous load alone. Experiments on two independent Kubernetes clusters, driven by real-world Azure Functions traces and repeated across multiple random seeds per cluster, quantify the gains. Relative to an otherwise identical single-layer LSTM ablation, our method reduces 90th-percentile latency by approximately 67\% and holds average latency within the 50\,ms hard SLO for 98.8\% of the run, against 49.6\% for that ablation and 43.5\% for the static HPA baseline. Against the KEDA reactive baseline, the gain is reliability rather than raw speed. Our agent reaches the almost same average latency at 75\% fewer replica-steps and 59\% less churn, with maximum P90 hard-SLO violation bursts of at most 5 consecutive intervals, against up to 24 for KEDA. Capacity sized from recent history therefore holds service levels with a smaller and steadier footprint, while capacity driven by instantaneous load defends the average and yields the sustained failures operators notice. Despite these gains, our study was limited to a single-tier microservice on controlled testbeds, abstracting away inter-service dependencies and the scale of multi-tenant production deployments.

Future research will extend the framework to a Multi-Agent Reinforcement Learning setting that coordinates scaling across dependent service chains, and to a large-scale bare-metal testbed where cross-node routing overheads and spatial load balancing become significant. Joint horizontal and vertical control, which adjusts CPU and memory limits alongside replica count, offers a further route to reducing resource stranding in heterogeneous edge nodes. We also plan to extend the evaluation beyond the CPU-bound workload studied here to I/O-bound and memory-bound serverless functions, where latency is shaped less by compute saturation and more by cold-start and initialization overheads. A broader baseline set, including predictive-HPA, ARIMA/Prophet-LSTM controllers, and other recent microservice autoscalers, would further contextualize the gains reported against KEDA. Finally, we aim to treat energy consumption as a first-class optimization objective and to validate the system on a physical 6G testbed.

\section*{Declaration of competing interest}
The authors declare that they have no known competing financial interests or personal relationships that could have appeared to influence the work reported in this paper.

\section*{Data Availability Statement}
The complete source code and experimental environment used in this study, along with the results, are available online~\citep{shaikh_github_2026}. The underlying workload data is sourced from the public Azure Functions traces~\citep{azurefunctions2021}.

\printcredits

\bibliographystyle{model1-num-names}
\bibliography{references}

@article{wang2023wireless,
  title={Wireless powered mobile edge computing networks: A survey},
  author={Wang, Xiaojie and others},
  journal={ACM Computing Surveys},
  volume={55},
  number={13s},
  pages={1--37},
  year={2023},
  publisher={ACM New York, NY}
}

@article{dogani2023auto,
  title={Auto-scaling techniques in container-based cloud and edge/fog computing: Taxonomy and survey},
  author={Dogani, Javad and Namvar, Reza and Khunjush, Farshad},
  journal={Computer Communications},
  volume={209},
  pages={120--150},
  year={2023},
  publisher={Elsevier}
}

@misc{shaikh_github_2026,
  author  = {Shaikh, Faraz},
  title   = {Stability-Aware Reinforcement Learning for Container Autoscaling: {GitHub} Repository},
  url     = {https://github.com/farazshaikh581/Autoscaling_mitigating-temporal-blindness},
  version = {1.0.0},
  year    = {2026},
}

@article{schulman2017proximal,
  title={Proximal policy optimization algorithms},
  author={Schulman, John and others},
  journal={arXiv preprint arXiv:1707.06347},
  year={2017}
}

@article{xu2025auto,
  title={Auto-scaling Approaches for Cloud-native Applications: A Survey and Taxonomy},
  author={Xu, Minxian and others},
  journal={arXiv preprint arXiv:2507.17128},
  year={2025}
}

@article{golec2024cold,
  title={Cold start latency in serverless computing: A systematic review, taxonomy, and future directions},
  author={Golec, Muhammed and others},
  journal={ACM Computing Surveys},
  volume={57},
  number={3},
  pages={1--36},
  year={2024},
  publisher={ACM New York, NY}
}

@misc{k8s_hpa_flapping,
  author       = {{The Kubernetes Authors}},
  title        = {{Horizontal Pod Autoscaling: Flapping}},
  howpublished = {Kubernetes Documentation},
  year         = {2025},
  url          = {https://kubernetes.io/docs/concepts/workloads/autoscaling/horizontal-pod-autoscale/#flapping},
  note         = {Accessed: 2025-12-28}
}

@inproceedings{rossi2020hierarchical,
  title={Hierarchical scaling of microservices in kubernetes},
  author={Rossi, Fabiana and Cardellini, Valeria and Presti, Francesco Lo},
  booktitle={2020 IEEE international conference on autonomic computing and self-organizing systems (ACSOS)},
  pages={28--37},
  year={2020},
  organization={IEEE}
}

@article{khaleq2021intelligent,
  title={Intelligent autoscaling of microservices in the cloud for real-time applications},
  author={Khaleq, Abeer Abdel and Ra, Ilkyeun},
  journal={IEEE access},
  volume={9},
  pages={35464--35476},
  year={2021},
  publisher={IEEE}
}

@inproceedings{xiao2022dscaler,
  title={Dscaler: A horizontal autoscaler of microservice based on deep reinforcement learning},
  author={Xiao, Zhijiao and Hu, Song},
  booktitle={2022 23rd Asia-Pacific Network Operations and Management Symposium (APNOMS)},
  pages={1--6},
  year={2022},
  organization={IEEE}
}

@article{10.1145/3465630,
author = {Sabuhi, Mikael and Mahmoudi, Nima and Khazaei, Hamzeh},
title = {Optimizing the Performance of Containerized Cloud Software Systems Using Adaptive PID Controllers},
year = {2021},
issue_date = {September 2020},
publisher = {Association for Computing Machinery},
address = {New York, NY, USA},
volume = {15},
number = {3},
doi = {10.1145/3465630},
journal = {ACM Trans. Auton. Adapt. Syst.},
month = aug,
articleno = {8},
numpages = {27}
}

@misc{hey,
  author       = {Jakub Rakyll},
  title        = {hey: HTTP load generator, ApacheBench (ab) replacement},
  year         = {2016},
  howpublished = {\url{https://github.com/rakyll/hey}},
  note      = {Accessed: 2026-02-16}
}

@misc{openfaas,
  author       = {OpenFaaS},
  title        = {OpenFaaS - Serverless Functions Made Simple},
  howpublished = {\url{https://www.openfaas.com/}},
  year = {2026},
  note         = {Accessed: 2026-02-16}
}

@misc{azurefunctions2021,
  author       = {{Microsoft Azure}},
  title        = {Azure Functions Invocation Trace 2021},
  year         = {2021},
  howpublished = {\url{https://github.com/Azure/AzurePublicDataset/blob/master/AzureFunctionsInvocationTrace2021.md}},
  note         = {Accessed: 2026-02-16}
}

@article{kim2022improved,
  title={Improved Q network auto-scaling in microservice architecture},
  author={Kim, Yeonggwang and others},
  journal={Applied Sciences},
  volume={12},
  number={3},
  pages={1206},
  year={2022},
  publisher={MDPI}
}

@article{lee2021deep,
  title={Deep Q-network-based auto scaling for service in a multi-access edge computing environment},
  author={Lee, Do-Young and others},
  journal={International Journal of Network Management},
  volume={31},
  number={6},
  pages={e2176},
  year={2021},
  publisher={Wiley Online Library}
}

@inproceedings{santos2025can,
  title={Can Reinforcement Learning be Generalized for Efficient Auto-Scaling in Containerized Clouds?},
  author={Santos, Jos{\'e} and others},
  booktitle={IEEE NOMS 2025},
  year={2025}
}

@inproceedings{gan2022adaptive,
  title={Adaptive auto-scaling in mobile edge computing: A deep reinforcement learning approach},
  author={Gan, Zhaoyu and Lin, Rongheng and Zou, Hua},
  booktitle={ICCECE 2022},
  pages={586--591},
  year={2022}
}

@article{panda2024faasctrl,
  title={FaaSCtrl: A Comprehensive-Latency Controller for Serverless Platforms},
  author={Panda, Abhisek and Sarangi, Smruti R},
  journal={IEEE Transactions on Cloud Computing},
  year={2024},
  publisher={IEEE}
}

@inproceedings{hausknecht2015deep,
  title={Deep Recurrent Q-Learning for Partially Observable MDPs.},
  author={Hausknecht, Matthew J and Stone, Peter},
  booktitle={AAAI fall symposia},
  volume={45},
  pages={141},
  year={2015}
}

@article{agarwal2024deep,
  title={A deep recurrent-reinforcement learning method for intelligent autoscaling of serverless functions},
  author={Agarwal, Siddharth and Rodriguez, Maria A and Buyya, Rajkumar},
  journal={IEEE Transactions on Services Computing},
  volume={17},
  number={5},
  pages={1899--1910},
  year={2024},
  publisher={IEEE}
}

@article{zhao2025mscnet,
  title={MSCNet: multi-scale network with convolutions for long-term cloud workload prediction},
  author={Zhao, Feiyu and others},
  journal={IEEE Transactions on Services Computing},
  year={2025},
  publisher={IEEE}
}

@inproceedings{zhang2025kis,
  title={KIS-S: A GPU-Aware Kubernetes Inference Simulator with RL-Based Auto-Scaling},
  author={Zhang, Guilin and others},
  booktitle={IPCCC2025},
  pages={1--8},
  year={2025},
  organization={IEEE}
}

@inproceedings{meybodi2022tedge,
  title={TEDGE-Caching: Transformer-based edge caching towards 6G networks},
  author={Meybodi, Zohreh Hajiakhondi and others},
  booktitle={IEEE ICC 2022},
  pages={613--618},
  year={2022}
}

@inproceedings{meng2023deepscaler,
  title={Deepscaler: Holistic autoscaling for microservices based on spatiotemporal gnn with adaptive graph learning},
  author={Meng, Chunyang and others},
  booktitle={2023 38th IEEE/ACM International Conference on Automated Software Engineering (ASE)},
  pages={53--65},
  year={2023},
  organization={IEEE}
}

@article{10.1145/2890784,
author = {Burns, Brendan and others},
title = {Borg, Omega, and Kubernetes},
year = {2016},
issue_date = {May 2016},
publisher = {Association for Computing Machinery},
address = {New York, NY, USA},
volume = {59},
number = {5},
issn = {0001-0782},
url = {https://doi.org/10.1145/2890784},
doi = {10.1145/2890784},
journal = {Commun. ACM},
month = apr,
pages = {50–57},
numpages = {8}
}

@INPROCEEDINGS{11297411,
  author={Shaikh, Faraz and Reali, Gianluca and Femminella, Mauro},
  booktitle={{CNSM} 2025}, 
  title={Intelligent Autoscaling with Attention-based Reinforcement Learning for SLA-Aware Resource Management in Edge-Cloud Environments}, 
  year={2025},
  volume={},
  number={},
  doi={10.23919/CNSM67658.2025.11297411}}

@article{santos2025gwydion,
  title={Gwydion: Efficient auto-scaling for complex containerized applications in Kubernetes through Reinforcement Learning},
  author={Santos, Jos{\'e} and others},
  journal={Journal of Network and Computer Applications},
  volume={234},
  pages={104067},
  year={2025},
  publisher={Elsevier}
}

@inproceedings{cortez2017resource,
  title={Resource central: Understanding and predicting workloads for improved resource management in large cloud platforms},
  author={Cortez, Eli and others},
  booktitle={Proceedings of the 26th Symposium on Operating Systems Principles},
  pages={153--167},
  year={2017}
}

@article{femminella2024application,
  title={Application of Proximal Policy Optimization for Resource Orchestration in Serverless Edge Computing.},
  author={Femminella, Mauro and Reali, Gianluca},
  journal={Computers (2073-431X)},
  volume={13},
  number={9},
  year={2024}
}

@article{ma2024auto,
  title={Auto-scaling and computation offloading in edge/cloud computing: a fuzzy Q-learning-based approach},
  author={Ma, Xiang and Zong, Kexuan and Rezaeipanah, Amin},
  journal={Wireless Networks},
  volume={30},
  number={2},
  pages={637--648},
  year={2024},
  publisher={Springer}
}

@article{gupta2025hybrid,
  title={A Hybrid Reactive-Proactive Auto-scaling Algorithm for SLA-Constrained Edge Computing},
  author={Gupta, Suhrid and Islam, Muhammed Tawfiqul and Buyya, Rajkumar},
  journal={arXiv preprint arXiv:2512.14290},
  year={2025}
}

@article{yan2021hansel,
  title={HANSEL: Adaptive horizontal scaling of microservices using Bi-LSTM},
  author={Yan, Ming and others},
  journal={Applied Soft Computing},
  volume={105},
  pages={107216},
  year={2021},
  publisher={Elsevier}
}

@article{guruge2025time,
  title={Time series forecasting-based kubernetes autoscaling using facebook prophet and long short-term memory},
  author={Guruge, Pasan Bhanu and Priyadarshana, YHPP},
  journal={Frontiers in Computer Science},
  volume={7},
  pages={1509165},
  year={2025},
  publisher={Frontiers Media SA}
}

@article{dimolitsas2026enabling,
  title={Enabling Multi-Application Multi-Objective Autoscaling With Quick Analytic Hierarchy Process},
  author={Dimolitsas, Ioannis and Dechouniotis, Dimitrios and Papavassiliou, Symeon},
  journal={IEEE Networking Letters},
  year={2026},
  publisher={IEEE}
}

@inproceedings{peng2023microservice,
  title={Microservice Auto-Scaling Algorithm Based on Workload Prediction in Cloud-Edge Collaboration Environment},
  author={Peng, Zijun and Tang, Bing and Xu, Wei and Yang, Qing and Hussaini, Ehsanullah and Xiao, Yuqiang and Li, Haiyan},
  booktitle={2023 IEEE International Conferences on Internet of Things (iThings) and IEEE Green Computing \& Communications (GreenCom) and IEEE Cyber, Physical \& Social Computing (CPSCom) and IEEE Smart Data (SmartData) and IEEE Congress on Cybermatics (Cybermatics)},
  pages={608--615},
  year={2023},
  organization={IEEE}
}

@misc{kubernetes_hpa,
  title        = {Horizontal Pod Autoscaling},
  author       = {{The Kubernetes Authors}},
  howpublished = {\url{https://kubernetes.io/docs/concepts/workloads/autoscaling/horizontal-pod-autoscale/}},
  year         = {2026},
  note         = {Accessed: 2026-01-15}
}

@article{punniyamoorthy2025slo,
  title={An SLO Driven and Cost-Aware Autoscaling Framework for Kubernetes},
  author={Punniyamoorthy, Vinoth and others},
  journal={arXiv preprint arXiv:2512.23415},
  year={2025}
}

@article{tran2024optimized,
  title={Optimized resource usage with hybrid auto-scaling system for knative serverless edge computing},
  author={Tran, Minh-Ngoc and Kim, YoungHan},
  journal={Future Generation Computer Systems},
  volume={152},
  pages={304--316},
  year={2024},
  publisher={Elsevier}
}

@inproceedings{hall2022opportunities,
  title={Opportunities for optimizing the container runtime},
  author={Hall, Adam and Ramachandran, Umakishore},
  booktitle={2022 IEEE/ACM 7th Symposium on Edge Computing (SEC)},
  pages={265--276},
  year={2022},
  organization={IEEE}
}

@article{park2023fully,
  title={Fully decentralized horizontal autoscaling for burst of load in fog computing},
  author={Park, EunChan and others},
  journal={Journal of Web Engineering},
  volume={22},
  number={6},
  pages={849--870},
  year={2023},
  publisher={River Publishers}
}

@article{kim2024self,
  title={Self-attention with temporal prior: can we learn more from the arrow of time?},
  author={Kim, Kyung Geun and Lee, Byeong Tak},
  journal={Frontiers in Artificial Intelligence},
  volume={7},
  pages={1397298},
  year={2024},
  publisher={Frontiers Media SA}
}

@article{lim2021temporal,
  title={Temporal fusion transformers for interpretable multi-horizon time series forecasting},
  author={Lim, Bryan and Ar{\i}k, Sercan {\"O} and Loeff, Nicolas and Pfister, Tomas},
  journal={International journal of forecasting},
  volume={37},
  number={4},
  pages={1748--1764},
  year={2021},
  publisher={Elsevier}
}

@misc{francois2025kubernetes,
  title        = {Kubernetes v1.33: HorizontalPodAutoscaler Configurable Tolerance},
  author       = {Fran\c{c}ois, Jean-Marc},
  year         = {2025},
  month        = apr,
  day          = {28},
  howpublished = {\url{https://kubernetes.io/blog/2025/04/28/kubernetes-v1-33-hpa-configurable-tolerance/}},
  note         = {Kubernetes Blog}
}

@article{hernandez2019survey,
  title={A survey and critique of multiagent deep reinforcement learning},
  author={Hernandez-Leal, Pablo and Kartal, Bilal and Taylor, Matthew E},
  journal={Autonomous Agents and Multi-Agent Systems},
  volume={33},
  number={6},
  pages={750--797},
  year={2019},
  publisher={Springer}
}

@article{BENEDETTI2026108112,
title = {Management of autoscaling serverless functions in edge computing via Q-Learning},
journal = {Future Generation Computer Systems},
volume = {175},
pages = {108112},
year = {2026},
issn = {0167-739X},
doi = {https://doi.org/10.1016/j.future.2025.108112},
author = {Priscilla Benedetti and Mauro Femminella and Gianluca Reali}
}

@ARTICLE{11430686,
  author={Mohajer, Amin and others},
  journal={IEEE Transactions on Network and Service Management}, 
  title={Joint SLA-Aware Task Offloading and Adaptive Service Orchestration With Graph-Attentive Multi-Agent Reinforcement Learning}, 
  year={2026},
  volume={23},
  number={},
  pages={3737-3754},
  doi={10.1109/TNSM.2026.3673188}}

@misc{keda,                                                                     title        = {{KEDA: Kubernetes Event-driven Autoscaling}},
    author       = {{CNCF KEDA Authors}},
    year = {2024},
    howpublished = {\url{https://keda.sh/}},note = {Accessed: 2026-08-01}
}

\end{document}